\documentstyle[prb,aps,multicol,epsfig,array]{revtex}
\begin{document}
\title{Incorporation of Density Matrix Wavefunctions in Monte Carlo 
Simulations: Application to the Frustrated Heisenberg Model}
\author{M. S. L. du Croo de Jongh, J. M. J. van Leeuwen and W. van Saarloos}
\address{Instituut--Lorentz, Leiden University,  P. O. Box 9506, 
2300 RA  Leiden, The Netherlands} 
\date{\today}
\maketitle

\begin{abstract}
We combine the Density Matrix Technique (DMRG) with Green 
Function Monte Carlo (GFMC) simulations.
Both methods aim to determine the groundstate of a quantum system but have
different limitations. The DMRG is most successful in 1-dimensional
systems and can only be extended to 2-dimensional systems 
for strips of limited width. GFMC is not restricted to low dimensions
but is limited by the efficiency of the sampling. This limitation is crucial
when the system exhibits a so--called sign problem, which on the other
hand is not a particular obstacle for the DMRG.
We show how to combine the virtues of both methods by using a DMRG
wavefunction as guiding wave function for the GFMC. This requires a
special representation of the DMRG wavefunction to make the simulations
possible within reasonable computational time. As a test case we apply
the method to the 2--dimensional frustrated Heisenberg antiferromagnet.
By supplementing the branching in GFMC with Stochastic Reconfiguration (SR) 
we get a stable simulation with a small variance also in the region
where the fluctuations due to minus sign problem are maximal. The 
sensitivity of the results to the choice of the guiding wavefunction is
extensively investigated.
 
We analyse the model as a function of the ratio of the next--nearest 
to nearest neighbor coupling strength which is a measure for the 
frustration. In agreement with earlier calculations it is found from 
the DMRG wavefunction that for small ratios
the system orders as a N\'eel type antiferromagnet and for large ratios
as a columnar antiferromagnet. The spin stiffness suggests an intermediate
regime without magnetic long range order. The energy curve indicates that 
the columnar phase is separated from the intermediate phase by a first 
order transition. The combination of DMRG and GFMC allows to substantiate
this picture by calculating also the spin correlations in the system.
We observe a pattern of the spin correlations in the intermediate
regime which is in--between dimerlike and plaquette type ordering,
states that have recently been suggested.
It is a state with strong dimerization in one direction and weaker 
dimerization in the perpendicular direction and thus it lacks the
the square symmetry of the plaquette state.

\end{abstract}
\vspace*{4mm}

PACS numbers: 75.40.Mg, 75.10.Jm, 02.70.Lq

\begin{multicols}{2}
\section{Introduction}

The Density Matrix Technique (DMRG) has proven to be a 
very efficient method to determine
the groundstate properties of low dimensional systems \cite{Wh1}. 
For a quantum chain it produces extremely accurate values for the 
energy and the correlation functions. In two
dimensional systems the calculational effort increases rapidly with the
size of the system. The most favorable geometry is that of a long small
strip. In practice the width of the strip is limited to around 8 to 10 lattice
sites. Greens Function Monte Carlo (GFMC)
is not directly limited by the size of the system but by
the efficiency of the importance sampling. When the system has a
minus sign problem the statistics is ruined in the long run and accurate
estimates are impossible. Many proposals \cite{sign} have been made to
alleviate or avoid the minus sign problem with varying success, but
all of them introduce uncontrollable errors in the sampling. In the DMRG
calculation of the wavefunction the minus sign problem is not manifestly
present. In all proposed cures of the minus sign problem the errors decrease
when the guiding wavefunction approaches the groundstate.

The idea of this paper is that DMRG wavefunctions are much better, also 
for larger systems, than the educated guesses which usually feature as
guiding wave functions. Moreover DMRG is a general
technique to construct a wavefunction without knowing too much about the
nature of the groundstate, with the possibility to systematically increase
the accuracy. Thus DMRG wavefunctions would do very well when they could be
used as guiding functions in the importance sampling of the GFMC.
There is a complicating factor which prevents a straightforward
implementation of this idea due to the fact that interesting systems
are so large that it is impossible to use a wavefunction via a look--up table.
The value of the wavefunction in a configuration has to be calculated by an
in--line algorithm. This has limited the guiding wavefunctions to simple 
expressions which are fast to evaluate. Consequently such guiding 
wavefunctions are not an accurate representation of the true 
groundstate wavefunction, in particular if the physics of the groundstate
is not well understood. In this paper we describe a method to read out
the DMRG wavefunction in an efficient way by using a 
special representation of the DMRG wavefunction. 

A second problem is 
that a good guiding wavefunction alleviates the minus sign problem, 
but cannot remove it as long as it is not exact. We resolve this dilemma
by applying the  method of Stochastic Reconfiguration which has recently 
been proposed by Sorella \cite{Sor}. The viability of our method is 
tested for the frustrated Heisenberg model.

The behavior of the Heisenberg antiferromagnet has been intruiging for a 
long time and still is in the center of research.
The groundstate of the antiferromagnetic 1-dimensional 
chain with nearest neighbor coupling is exactly known. In higher dimensions 
only approximate theories or simulation results are available. The source of 
the complexity of the groundstate are the large quantum fluctuations
which counteract the tendency of classical ordering. The unfrustrated
2--dimensional Heisenberg antiferromagnet orders in a N\'eel state and
by numerical methods the properties of this state can be analyzed accurately
\cite{San}. The situation is worse when the interactions are 
competing as in a 2-dimensional square lattice with antiferromagnetic 
nearest neighbor $J_1$ and next nearest neighbor $J_2$ coupling. 
This spin system with
continuous symmetry can order in 2 dimensions at zero temperature, but 
it is clear that the magnetic order is frustrated by the opposing 
tendencies of the two types of interaction. The ratio $J_2/J_1$ is a 
convenient parameter for the frustration. For small values the system 
orders antiferromagnetically in a N\'eel type arrangement, which 
accomodates the nearest neighbor interaction. For large ratios a magnetic 
order in alternating columns of aligned spins (columnar phase) will prevail; 
in this regime the roles of the two couplings are reversed: the nearest
neighbor interaction frustrates the order imposed by the next nearest neighbor
interaction. In between, for ratios of the order of 0.5, the frustration is
maximal and it is not clear which sort of groundstate results. This problem
has been attacked by various methods but not yet by DMRG and
only very recently by  GFMC \cite{Sor2}.
This paper addresses the issue by studying the spin correlations.

A simple road to the answer is not possible since the behavior of 
the system with frustration presents some fundamental problems. 
The most severe obstacle is that frustration implies a sign problem which
prevents the straightforward use of the GFMC simulation technique. 
Moreover the frustration substantially complicates the structure of the 
groundstate wavefunction. Generally frustration encourages the formation 
of local structures such as dimers and plaquettes which are at odds, but not
incompatible, with long range magnetic order. These correlation patterns
are the most interesting part of the intermediate phase and the main
goal of this investigation. 

Many attempts have been made to clarify the situation. Often 
simple approximations  such as mean--field or spin--wave theory
give useful information about the qualitative behavior of the phase diagram.
A fairly sophistocated mean--field theory using the Schwinger boson 
representation does not give an intermediate phase \cite{Duc}.
Given the complexity of the phase diagram and the subtlety of the effects
it is not clear whether such approximate methods can give in this case 
a reliable clue to the qualitative behavior of the system. 

Exact calculations have been performed on small systems up to 
size $6 \times 6$ by Schulz et al. \cite{Sch}.
Although this information is very accurate and unbiased to possible phases,
the extrapolation to larger systems is a long way, the more so 
in view of indications
that the anticipated finite size behavior only applies for larger systems. 
Another drawback of these small systems is that the groundstate is assumed
to have the full symmetry of the lattice. Therefore the symmetry
breaking, associated with the formation of dimers, ladders or plaquettes,
which is typical for the intermediate state, can not be observed directly.

More convincing are the systematic series expansion as reported 
recently by Kotov et al. \cite{Kot},\cite{Kot2} and by Singh et al. 
\cite{Sin}, which bear on an infinite system. They start with an independent 
dimers (plaquettes) and study the series expansion in the coupling 
between the dimers (plaquettes). By the choice of the state, around which
the perturbation expansion is made, the type of spatial symmetry breaking
is fixed. These studies favor in the intermediate
regime the dimer state over the plaquette state. Their dimer state has 
dimers organized in ladders in which the
chains and the rungs have nearly equal strength. So the system breaks the 
translational invariance only in one direction. The energy differences
are however small and the series is finite, so further investigation is
useful. Our simulations yield correlations in good agreement with theirs,
but do not confirm the picture of translational invariant ladders. Instead
we find an additional weaker symmetry breaking {\it along} the ladders, such
that we come closer to the plaquette picture.

Very recently Capriotti and Sorella \cite{Sor2} have carried out a GFMC 
simulation for $J_2 = 0.5J_1$ and have studied the susceptibilities for
the orientational and translational symmetry breaking. They conclude
that the groundstate is a plaquette state with full
symmetry between the horizontal and vertical direction.

From the purely theoretical side the problem has been discussed by Sachdev and
Read \cite{Sac} on the basis of a large spin expansion. From their analysis a
scenario emerges in which the N\'eel phase disappears upon increasing
frustration in a continuous way. Then a gapped spatial--inhomogeneous
phase with dimerlike correlations appears. For even higher 
frustration ratios a first order transition takes place to the columnar phase.
Although this scenario is qualitative, without precise location of the
phase transition points, it definitively excludes dimer formation
in the magnetically ordered N\'eel and columnar phase. It is remarkable
that two quite different order parameters (the magnetic order and the
dimer order) disappear simultaneously and continuously on opposite sides 
of the phase transition. In this scenario, this is taken as an indication of
some kind of duality of the two phases. 
 
Given all these predictions it is of utmost interest to further study the 
nature of the intermediate state. Due to the smallness of the differences in 
energy between the various possibilities, the energy will not be the ideal 
test for the phase diagram. Therefore we have decided to focus directly 
on the spin correlations as a function of the ratio $J_2/J_1$.
In this paper we first investigate the 2--dimensional frustrated Heisenberg 
model by constructing the DMRG wave function of the groundstate for long 
strips up to a width of 8 sites.
The groundstate energy and the spin stiffness which are calculated,
confirm the overal picture described above, but the results are not
accurate enough to allow for a conclusive extrapolation to larger systems.
Then we study an open 10x10 lattice by means of the GFMC technique using
DMRG wavefunctions as guiding wavefunction for the importance sampling.
The GFMC are supplemented with Stochastic Reconfiguration as proposed
by Sorella \cite{Sor} as an extension of the Fixed Node technique \cite{Cep}.
This method avoids
the minus sign problem by replacing the walkers regularly by a new set
of positive sign with the same statistical properties. The first observation
is that GFMC improves the energy of the DMRG in a substantial and systematic
way as can be tested in the unfrustrated model where sufficient information
is available from different sources. Secondly the spin correlations 
become more accurate and less dependent on the technique used for constructing
the DMRG wavefunction. The DMRG technique is focussed on the energy of 
the system and less on the correlations. The GFMC probes mostly the local 
correlations of the system as all the moves are small and correspond to local
changes of the configurations. With these spin correlations we investigate
the phase diagram for various values of the frustration ratio $J_2/J_1$.

The paper begins with the definition of the model to avoid ambiguities. Then
a short description of our implementation of
the DMRG method is given. We go into more detail
about the way how the constructed wavefunctions can be used as guiding
wavefunctions in the GFMC simulation. This is a delicate problem since the full
construction of a DMRG wavefunction takes several hours on a workstation.
Therefore we separate off the construction of the wavefunction and cast it in
a form where the configurations can be obtained from each other by matrix
operations on a vector. So the length of the computation of the wavefunction
in a configuration scales with the square of the number of states 
included in the DMRG wavefunction. But even then the actual 
construction of the value of the wavefunction in a given configuration 
is so time consuming that utmost effiency must be reached in obtaining
the wavefunction for successive configurations. The remaining
sections are used to outline the GFMC and the Stochastic Reconfiguration
and to discuss the results. We concentrate on the correlation
functions since we see them as most significant for the structure of the
phases. We give first a global evaluation of the correlation function patterns 
for a wide set of frustration ratios and then focus on a number of points 
to see the dependence on the guiding wavefunction and to deduce the trends.
The paper closes with a discussion and a comparison with other results  
in the literature.

\section{The Hamiltonian}

The hamiltonian of the system refers to spins on a square lattice.
\begin{equation} \label{a1}
{\cal H} = J_1 \sum_{(i,j)} {\bf S}_i \cdot {\bf S}_j + J_2
\sum_{[i,j]} {\bf S}_i \cdot {\bf S}_j. 
\end{equation} 
The ${\bf S}_i$ are spin $\frac{1}{2}$ operators and 
the sum is over pairs of nearest neigbors $(i,j)$ and over pairs of
next nearest neighbors $[i,j]$ on a quadratic lattice. Both coupling 
constants $J_1$ and $J_2$ are supposed to be positive.
$J_1$ tries to align the nearest neigbor spin in an 
antiferromagnetic way and $J_2$ tries to do the same with the next nearest
neighbors. So the spin system is frustrated, implying an intrinsic minus 
sign in the simulations that cannot be gauged away by a rotation of the
spin operators.
\begin{figure}
    \centering \epsfxsize=6cm
    \epsffile{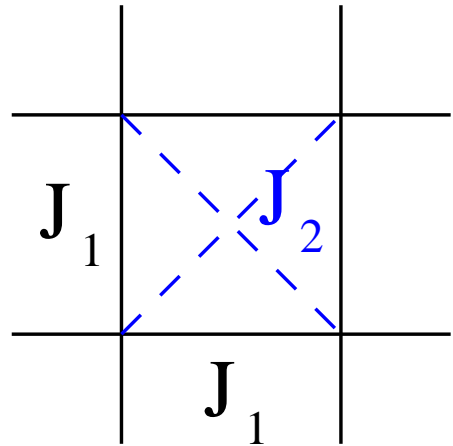}
   \caption{The interaction constants $J_1$ and $J_2$}
    \label{fig0}
\end{figure}

In order to prepare for the representation of the hamiltonian we express
the spin components in spin raising and lowering operators
\begin{equation} \label{a2}
{\bf S}_i \cdot {\bf S}_j = \frac{1}{2} (S^+_i S^-_j + S^-_i S^+_j)
  + S^z_i S^z_j.
\end{equation}
We will use the $z$ component representation of the spins and a complete
state of the spins will be represented as
\begin{equation} \label{a3}
| R \rangle = |s_1, s_2, \cdots , s_N \rangle,
\end{equation}
where the $s_j$ are eigenvalues of the $S^z_j$ operator. 
The diagonal matrix elements of the hamiltonian 
are in the representation (\ref{a3}) given by
\begin{equation} \label{a4}
\langle R | {\cal H} | R \rangle =J_1 \sum_{(i,j)} s_i s_j + 
J_2 \sum_{[i,j]} s_i s_j. 
\end{equation}
The off-diagonal elements are between two nearby configurations 
$R'$ and $R$. $R'$ is the same as $R$ except at a pair of nearest 
neighbors sites $(i,j)$ or next nearest neighbor sites $[i,j]$, for
which the spins $s_i$ and $s_j$ are opposite. In $ R'$ the pair is turned
over by the hamiltonian. Then
\begin{equation} \label{a5}
\langle R' | {\cal H} | R \rangle = \frac{1}{2} J_1 \hspace*{1cm}
{\rm or} \hspace*{1cm} \langle R' | {\cal H} | R \rangle = \frac{1}{2} J_2,
\end{equation}
depending on whether a nearest or a next nearest pair is flipped.

\section{The DMRG Procedure}

The DMRG procedure approximates the groundstate wavefunction by searching
through various representations in bases of a given dimension $m$ \cite{Wh1}.
Here we take the standard method (with two connecting sites) for granted
and make the preparations for the extraction of the wavefunction.
\begin{figure}[h]
   \centering \epsfxsize=\linewidth
   \epsffile{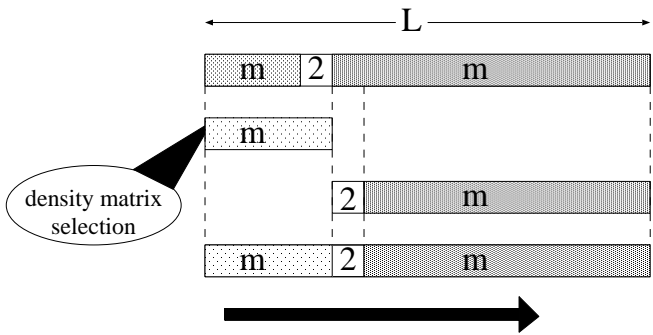}
   \caption{The DMRG procedure with one connecting site}
   \label{fig1}
\end{figure}

The system is mapped onto a 1-dimensional chain (see  Fig. \ref{fig1}) and
separated into two parts: a {\it left} and {\it right}  hand part.
They are connected by one site. Each part is represented in
a basis of at most $m$ states. With a representation of all the operators
in the hamiltonian in these bases one can find the groundstate of the
system. We thus have several representations of the groundstate
depending on the way in which the system is divided up into subsystems. 
The point is to see how these representations are connected and
how they possibly can be improved. We take a representation for the
right hand parts and improve those on the left.
So we assume that for a given division we have the groundstate of the
whole system and we want to
enlarge the left hand side at the expense of the right hand side. The
first step is to include the connecting site in the left hand part. This
enlarges the basis for the left hand side from $m$ to $2m$ and a selection
has to be made of $m$ basis states. This goes with the help of the density
matrix for the left hand side as induced by the wavefunction for the whole
system. For later use we write out the basic equations for the density
matrix in the configuration representation. Let, at a certain stage in the
computation, $| \Phi \rangle$ be the approximation to the groundstate.
The configurations of the right hand part and the left hand part are 
denoted by  $R_r$ and $R_l$.
Then the density matrix for the left hand part reads
\begin{equation} \label{b1}
\langle R_l |\rho | R'_l \rangle = \sum_{R_r} \langle R_l, R_r | \Phi \rangle
                      \langle \Phi | R'_l, R_r \rangle.
\end{equation}
In practice we do not solve the eigenvalues of the density matrix in
the configuration representation, but in a projection on a smaller basis. 
White \cite{Wh1}  has shown that the best way to represent the state 
$| \Phi \rangle$ is to select the $m$ eigenstates $| \alpha \rangle$ 
with the largest eigenvalue
\begin{equation} \label{b2}
\sum_{R'_l} \langle R_l |\rho | R'_l \rangle \langle R'_l | \alpha \rangle =
\lambda_\alpha \langle R_l | \alpha \rangle.
\end{equation}

The next step is to break up the right hand part into a connecting
site and a remainder. With the basis for this remainder and the newly
acquired basis for the left hand part we can again compute the groundstate 
of the whole system as indicated in the lower part of the figure. Now we are 
in the same position as we started, with the difference that the connecting
site has moved one position to the right. Thus we may repeat the cycle
till the right hand part is so small that it can exactly be represented
by $m$ states or less. Then we have constructed for the left hand part
a new set of bases, all containing $m$ states, for system parts of variable 
length. Next we reverse the roles of {\it left} and {\it right} 
and move back in order to improve the bases for the right hand parts 
with the just constructed bases for the left hand part.

The process may be iterated till it converges towards a steady state. The 
great virtue of the method is that it is variational. In each step the
energy will lower till it saturates. In 1-dimensional system the method
has proven to be very accurate \cite{Wh1}. So one wonders what the 
main trouble is in higher dimensions. 
\begin{figure}[h]
   \centering \epsfxsize=\linewidth
   \epsffile{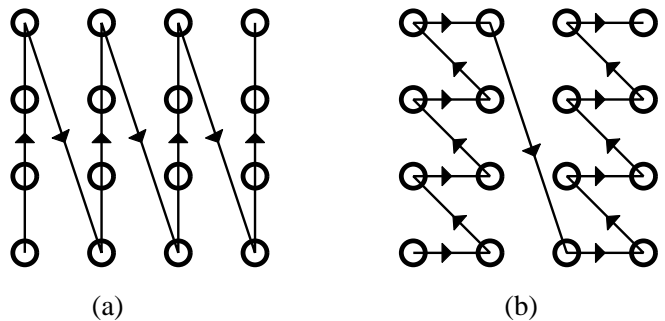}
   \caption{Two 1--dimensional paths through the system: ``straight'' (a)
            and ``meandering'' (b).}
   \label{fig2}
\end{figure}

In Fig. \ref{fig2} we have drawn 2 possible ways to map the
system on a 1-dimensional chain. One sees that if we divide again 
the chain into a left hand part and a right hand part and a connecting
site, quite a few sites of the left hand part
are nearest or next nearest neighbors of sites of the right hand part. So 
the coupling between the two parts of the chain is not only through the
connecting site but also through sites which are relatively far away from
each other in the 1-dimensional path. The operators for the spins on these
sites are not as well represented as those of the connecting site, which 
is fully represented by the two possible spin states. Yet the correlations
between the interacting sites count as much for the energy of the system
as those interacting with the connecting site. One may say that the further
away two interacting sites are in the 1-dimensional chain the poorer their
influence is accounted for. This consideration explains in part why open
systems can be calculated more accurately than closed systems, even in 
1-dimensional systems.

It is an open question which map of the 2-dimensional onto a 1-dimensional
chain gives the best representation of the groundstate of the system. 
Also other divisions of the system
than those suggested by a map on a 1--dimensional chain are possible and 
we have been experimenting with arrangements which reflect better the
2--dimensional character of the lattice\cite{Luc}. They are
promising but the software for these is not as sophisticated as the 
one developed by White \cite{Wh2} for the 1-dimensional chain. 
We therefore have restricted our calculations to the two paths shown here. 
The second choice. the ``meandering'' path, was motivated
by the fact that it has the strongest correlated sites most nearby in the
chain and this choice was indeed justified by a lower energy for a
given dimension $m$ of the representation than for the ``straight'' path.

The DMRG calculations as well as the corresponding GFMC simulations 
are carried out for both paths. The meandering path has to be
preferred over the straight path as the DMRG wavefunctions 
generally give a better energy value and the simulations suffer less from 
fluctuations. Nevertheless we have also investigated the straight path,
since the path chosen leaves its imprints on the resulting
correlation pattern and the paths break the symmetries in different ways.
Both paths have an orientational preference. In open systems the 
translational symmetry is broken anyway, but the meandering path
has in addition a staggering in the horizontal direction. This together
with the horizontal nearest neighbor sites appearing in the meandering 
path gives a preference for horizontal dimerlike correlations in this path.
On the other hand the straight path prefers the dimers in the vertical
direction. Comparing the results of the two choices, allows us to draw further 
conclusions on the nature of the intermediate state.

\section{Extracting configurations from the DMRG wavefunction}

It is clear that the wavefunction which results from a DMRG-procedure is
quite involved and it is not simple to extract its value for a given
configuration.  We assume now that the DMRG wavefunction
has been obtained by some procedure and we will give below an 
algorithm to obtain efficiently the value for an arbitrary configuration
(see also \cite{Luc} for an alternative description).

The first step is the construction of a set of representations for the
wavefunction in terms of two parts (without a connecting site in between). 
Let the left hand part contain $l$ sites and the other part $N-l$ sites.
We denote the $m$ basis states of the left hand part by the index $\alpha$
and those of the right hand part by $\bar{\alpha}$.
The eigenstates of the two parts are closely linked and related as follows
\begin{equation} \label{c1}
\left\{ \begin{array}{rcl}
\langle R_l | \alpha  \rangle & = & \displaystyle
\frac{1}{\sqrt{\lambda_\alpha}}   \sum_{R_r} \langle \Phi |  R_l, R_r  \rangle 
\langle R_r | \bar{\alpha} \rangle,  \\*[3mm]
\langle R_r | \bar{\alpha}  \rangle & = & \displaystyle 
\frac{1}{\sqrt{\lambda_\alpha}}  \sum_{R_l} 
\langle \alpha | R_l \rangle \langle R_l, R_r | \Phi \rangle. 
\end{array} \right.
\end{equation}
It means that for every eigenvalue $\lambda_\alpha$ there is and eigenstate
$\alpha$ for the left hand part and an $\bar{\alpha}$ for the right hand
density matrix. The proof of (\ref{c1}) follows from insertion in the
density matrix eigenvalue equation (\ref{b2}).

The second step is a relation for the groundstate wavefunction in terms of
these eigenfunctions. Generally we have
\begin{equation} \label{c2}
\langle R_l, R_r | \Phi \rangle = \sum_{\alpha, \bar{\beta}}
\langle R_l | \alpha \rangle \langle R_r | \bar{\beta} \rangle 
\langle \alpha \bar{\beta} | \Phi \rangle, 
\end{equation}
while due to (\ref{c1}) we find
\begin{equation} \label{c3}
\begin{array}{rcl}
\langle \alpha \bar{\beta} | \Phi \rangle & = & \displaystyle 
\sum_{R_l, R_r} \langle \alpha |
R_l \rangle \langle \bar{\beta} | R_r \rangle \langle R_l, R_r | \Phi \rangle
\\*[2mm] 
 & = &  \displaystyle\sqrt{\lambda_\alpha} \sum_{R_r} \langle \bar{\beta} 
| R_r \rangle \langle R_r | \bar{\alpha} \rangle \ = \delta_{\alpha, \beta} 
\sqrt{\lambda_\alpha}.
\end{array}
\end{equation}
Thus we can represent the groundstate as
\begin{equation} \label{c4}
\langle R_l, R_r | \Phi \rangle = \sum_\alpha \sqrt{\lambda^l_\alpha} 
\langle R_l |\alpha \rangle_l \langle R_r | \bar{\alpha} \rangle_{N-l}.
\end{equation}
For this part of the problem we have to compute and store the set 
of $m$ eigenvalues $\lambda^l_\alpha$ for each division $l$.
We point out again that we have on the left hand side the wavefunction
and on the right hand side representations for given division $l$,
which all lead to the same wavefunction.
The last step is to see the connection between these representations.

As intermediary we consider a representation of the wavefunction with
one site $s_l$ separating the spins $s_1 \cdots s_{l-1}$ on the left hand side
from $s_{l+1} \cdots s_N$ on the right hand side. Using the same basis as
in (\ref{c4}) we have
\begin{equation} \label{c5}
\begin{array}{l}
\langle s_1 \cdots s_{l-1}, s_l, s_{l+1} \cdots s_N | \Phi \rangle = \\*[2mm]
\sum_{\alpha, \alpha'} \langle s_1 \cdots s_{l-1} | \alpha \rangle
\phi^l_{\alpha,\alpha'} (s_l) \langle s_{l+1} \cdots s_N |\bar{\alpha'}\rangle.
\end{array}
\end{equation}
We compare this representation in two ways with (\ref{c4}). First we
combine the middle site with the left hand part. This leads to $m$ states
which can be expressed as linear combinations of the states of the 
enlarged segment
\begin{equation} \label{c6}
\sum_\alpha \langle s_1 \cdots s_{l-1} | \alpha \rangle 
\phi^l_{\alpha,\alpha'} (s_l) = \sum_{\alpha''} 
\langle s_1 \cdots s_l | \alpha'' \rangle T^l_{\alpha'', \alpha'}.
\end{equation}
In fact this relation is the very essence of the DMRG procedure. The 
wave function in the larger space is projected on the eigenstates of the
the density matrix of that space. Since the process of zipping back 
forth has converged there is indeed a fixed
relation (\ref{c6}). However when we insert (\ref{c6}) into (\ref{c5}) and
compare it with (\ref{c4}) we conclude that the matrix $T$ must be diagonal
\begin{equation} \label{c7}
 T^l_{\alpha'', \alpha'}  = \delta_{\alpha'',\alpha'} 
\sqrt{\lambda^l_{\alpha'}}.
\end{equation}
This leads to the recursion relation
\begin{equation} \label{c8}
\langle s_1 \cdots s_l | \alpha' \rangle = \sum_{\alpha} 
   \langle s_1  \cdots s_{l-1} | \alpha \rangle A^l_{\alpha,\alpha'}(s_l)
\end{equation}
with 
\begin{equation} \label{c9}
 A^l_{\alpha,\alpha'}(s_l) = \phi^l_{\alpha,\alpha'} (s_l)/
\sqrt{\lambda^l_{\alpha'}}.
\end{equation}
The second combination concerns the contraction of the middle site with the
right hand part. This leads to the recursion relation
\begin{equation} \label{c10}
\langle s_l \cdots s_N | \bar{\alpha} \rangle = \sum_{\alpha'} 
  B^{l-1}_{\alpha, \alpha'}(s_l) \langle s_{l+1}  \cdots s_N | \bar{\alpha}' 
\rangle
\end{equation}
with
\begin{equation} \label{c11}
B^{l-1}_{\alpha, \alpha'}(s) = \phi^l_{\alpha,\alpha'} (s)/
\sqrt{\lambda^{l-1}_\alpha} .
\end{equation}
The $A$ and $B$ matrices are the essential ingredients of the
calculation of the wavefunction. From (\ref{c11}) and (\ref{c9}) follows that
they are related as
\begin{equation} \label{c12}
B^{l-1}_{\alpha, \alpha'} (s) =\sqrt{\lambda^l_{\alpha'}\, /\,
\lambda^{l-1}_\alpha} \,A^l_{\alpha,\alpha'} (s).
\end{equation}

By the recursion relations the basis states are expressed as products of
$m \times m$ matrices. The determination of the DMRG wavefunction and
the matrices $A$ (or $B$) is part of the determination of the DMRG
wavefunction which is indeed lengthy but fortunately no part of the 
simulation. The matrices can be stored and contain the information
to calculate the wavefunction for any configuration. The value of the 
wavefunction is now obtained as the product of matrices acting on a
vector. Thus the calculational effort scales with $m^2$. Using  
relation (\ref{c12}) one reconfirms by direct calculation that the 
wavefunction is indeed independent of the division $l$.

When the simulation is in the  
configuration $R$, all the $\langle R_l | \alpha \rangle_l$ and the 
$\langle R_r | \bar{\alpha} \rangle_{N-l}$ are calculated and stored, with
the purpose to calculate the wavefunctions more efficiently for the
configurations $R'$ which are connected to $R$ by the hamiltonian and which
are the candidates for a move. The structure of of these nearby states
is $R' = s_1 \cdots s_{l_2} \cdots s_{l_1} \cdots s_N \quad (l_2 > l_1)$. 
So we have that for $R'$ the representation
\begin{equation} \label{c13}
\langle R' | \Phi \rangle = \sum_\alpha \sqrt{\lambda^{l_2}_\alpha}
\langle s_1 \cdots s_{l_2} \cdots s_{l_1} | \alpha \rangle
\langle s_{l_2 + 1} \cdots s_N | \bar{\alpha} \rangle
\end{equation}
holds. Now we see the advantage of having the wavefunction stored for all the
divisions. The second factor in (\ref{c13}) is already tabulated; 
the first factor involves a number
of matrix multiplications equal to the distance in the chain of the two
spins $l_1$ and $l_2$ till one reaches a tabulated function. One can use
the tables for a certain number of moves but after a while it starts to pay
off to make a fresh list. 

\section{Green Function Monte Carlo simulations}

The GFMC technique employs the operator
\begin{equation} \label{d1}
{\cal G} = 1 - \epsilon {\cal H}
\end{equation}
and uses the fact that the groundstate $| \Psi_0 \rangle$ results from 
\begin{equation} \label{d2} 
| \Psi_0 \rangle \sim {\cal G}^n | \Phi \rangle, \quad \quad \quad
 \epsilon \ll 1, \quad \quad \quad n \epsilon \gg 1
\end{equation}
where in principle $| \Phi \rangle $ may be any function which is 
non-orthogonal to the groundstate. In view of the possible symmetry breaking,
the overlap is a point of serious concern on which we come back in the 
discussion. In practice we will use the best
$| \Phi \rangle $ that we can construct conveniently by the DMRG--procedure
described above. The closer $| \Phi \rangle $ is to the groundstate the 
smaller the number of factors $n$ in the product needs to be in order 
to find the groundstate. 
Evaluating (\ref{d2}) in the spin representation gives for the projection on 
the trial wavefunction the following long product
\begin{equation} \label{d3}
\langle \Phi|\Psi_0 \rangle  \simeq \sum_{\bf R} \langle \Phi |R_M \rangle 
\left[ \prod^M_{i=1}
\langle R_i | {\cal G} |R_{i-1} \rangle \right] \langle R_0 | \Phi \rangle.
\end{equation}
Here the sum is over paths ${\bf R} = (R_M, \cdots R_1,R_0)$ which will be
generated by a Markov process.
The Markov process involves a transition probability 
$T(R_i \leftarrow R_{i-1})$ and the averaging process uses a 
weight $m(R)$. Its is natural to connect the
transition probabilities to the matrix elements of the Greens Function 
${\cal G}$. But here comes the sign problem into the game: 
the transition probabilities have to be positive (and normalized). 
So we put the transition rate proportional
to the absolute value of the matrix element of the Greens Function
\begin{equation} \label{d4}
T(R \leftarrow R') = \frac{|\langle R |{\cal G} | R'\rangle|}
{\sum_{R''} | \langle R'' | {\cal G} | R' \rangle |}.
\end{equation}
This implies that we have to use  a sign function $s(R,R')$
\begin{equation} \label{d5}
s(R,R') = \frac{\langle R |{\cal G} | R'\rangle}
               {|\langle R |{\cal G} | R'\rangle|}
\end{equation}
and a weight factor
\begin{equation} \label{d6}
m(R) = \sum_{R'} | \langle R' | {\cal G} | R \rangle |.
\end{equation}
All these factors together form the matrixelement of the Green Function
\begin{equation} \label{d7}
\langle R |{\cal G} | R'\rangle = T(R \leftarrow R') s(R,R') m(R').
\end{equation}
If the matrix elements of the Greens Function were all positive, or could
be made positive by a suitable transformation,
we would not have to introduce the sign function. We leave
its consequences to the next section.
By the representation (\ref{d7}) we can write the contribution of the path
as a product of transition probabilities, signs and local weights. The 
transition probabilities control the growth of the
Markov chain. The signs and weights constitute the weight of a path
\begin{equation} \label{d8}
M({\bf R}) = m_f (R_M) \left[ \prod^M_{i=1} s(R_i, R_{i-1}) m(R_{i-1})\right]
m_i (R_0).
\end{equation}
The initial and final weight have to be chosen such that the weight of the
paths corresponds to the expansion (\ref{d3}). For the innerproduct
$\langle \Phi | \Psi_0 \rangle$ we get
\begin{equation} \label{d9}
m_i (R) = \langle R | \Phi \rangle, \quad \quad \quad
m_f (R) = \langle \Phi | R \rangle.
\end{equation}
With this final weight we have projected the groundstate on the trial
wave. This allows us to calculate the so--called mixed averages.
For that purpose we define the local estimator ${\cal O}$
\begin{equation} \label{d10}
O(R) = \frac{\langle \Phi | {\cal O} | R \rangle}
{\langle \Phi | R \rangle}\, ,
\end{equation}
which yields the mixed average
\begin{equation} \label{d11}
\langle {\cal O} \rangle_m \equiv \frac{\langle \Phi |{\cal O} | 
\Psi_0  \rangle}{\langle \Phi | \Psi_0 \rangle} =
\frac{\sum_{{\bf R}} O(R_M) M({\bf R})}{\sum_{{\bf R}} M({\bf R})}.
\end{equation}
For operators not commuting with the hamiltonian the mixed average is
an approximation to the groundstate average. Later on we will improve on it.

In this raw form the GFMC would hardly work because all paths are generated
with equal weight. One can do better by importance sampling in which one
transforms the problem to a Greens Function with matrix elements
\begin{equation} \label{d12}
\langle R | \tilde{{\cal G}} | R' \rangle = \frac{\langle \Phi | R \rangle
\langle R | {\cal G} | R' \rangle}{ \langle \Phi | R' \rangle}.
\end{equation}
Generally this can be seen as a similarity transformation on the 
operators and from now on
everywhere operators with a tilde are related to their counterpart without a 
tilde as in (\ref{d12}). It gives only a minor change in the formulation.
The transition rates are based on the matrix elements of $\tilde{{\cal G}}$
and so are the signs and weights. Thus we have a set of definitions like
(\ref{d4})--(\ref{d6}) with everywhere a tilde on top. It leads also to 
a change of the initial and final weight
\begin{equation} \label{d13}
\tilde{m}_i (R) = |\langle \Phi | R \rangle|^2, \quad \quad \quad 
\tilde{m}_f (R) = 1.
\end{equation}
By chosing these weights the formula (\ref{d10}) for the average still
applies with a weight $\tilde{M} ({\bf R})$ made up as in (\ref{d8}) with
the weights and signs with a tilde.
Using the tilde operators the local estimator (\ref{d10}) reads
\begin{equation} \label{d14}
O(R) = \sum_{R'} \langle R' |\tilde{{\cal O}} | R \rangle.
\end{equation}

We will speak about the various paths in terms of independent walkers
that sample these paths.
As some walkers become more important than others in the process, it is
wise to improve the variance by branching, which we will discuss later with
the sign problem. Before we embark on the discussion of the
sign problem we want to summarize a number of aspects of the GFMC simulation
relevant to our work.
\begin{itemize}
\item The steps in the Markov process are small, only the ones induced
by one term of the hamiltonian feature in a transition to a new state. This 
makes the subsequent states quite correlated. So many steps have to be
performed before a statistical independent configuration is reached; on the
average a number of the order of the number of sites.
\item In every configuration the wave function for a number of neighboring 
states (the ones which are reachable by the Greens Function), has to be
evaluated. This is a time consuming operation and it makes the simulation
quite slow, because out of the possibilities (of order $N$) only one
is chosen and all the information gathered on the others is virtually useless.
\item The necessity to choose a small $\epsilon$ in the Greens Function seems
a further slow down of the method, but it can be avoided by the technique
of continuous time steps developed by Ceperley and Trivedi \cite{Tri}.
In this method the
possibility of staying in the same configuration (the diagonal element of the
Greens Function) is eliminated and replaced by a waiting time before a move
to another state is made. (For further details in relation to the present 
paper we refer to \cite{Luc})
\item The average (\ref{d10}) can be improved by replacing it by
\begin{equation} \label{d15}
\langle {\cal O} \rangle_{im} = 2 \frac{\langle \Phi | {\cal O} | \Psi_0
\rangle}{\langle \Phi | \Psi_0 \rangle} - \frac{\langle \Phi | {\cal O}
| \Phi \rangle}{\langle \Phi | \Psi_0  \rangle}
\end{equation}
of which the error with respect to the true average is of second order in
the deviation of $|\Phi \rangle$ from $|\Psi_0 \rangle$. For conserved
operators, such as the energy, this correction is not needed since the
the mixed average gives already the correct value.
\end{itemize}

\section{The sign problem and its remedies}

In the hamiltonian (\ref{a1}) the $z$ component of the spin operator keeps
the spin configuration invariant, whereas the $x$ and $y$ components change
the configuration. The typical change is that a pair of nearest or next
nearest neighbors is spin reversed. Inspecting the Greens Function it means
that all changes to another configuration involve a minus sign! Thus the
Greens Function is as far as possible from the ideal of positive matrix 
elements. The diagonal terms are positive, but they always are positive for 
sufficiently small $\epsilon$. Importance sampling can remove minus signs
in the transition rates, when the ratio of the guiding wavefunction involves
also a minus sign. For $J_2 \neq 0$ no guiding wavefunction can remove the
minus sign problem completely. In Fig. 3 we show a loop of two nearest
neighbor spin flips followed by a flip in a next nearest neighbor pair, 
such that the starting configuration is restored. The product of
the ratios of the guiding wavefunction drops out in this loop, but the
product of the three matrix elements has a minus sign. So at least 
one of the transitions must involve a minus sign.
\begin{figure}[h]
   \centering \epsfxsize=\linewidth
   \epsffile{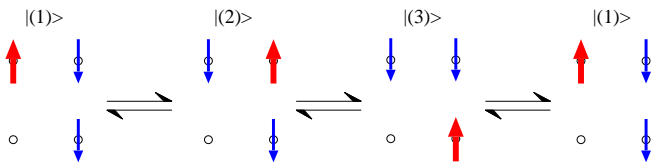}
   \caption{Illustration of the sign problem in the frustrated Heisenberg 
model. The shown sequence of spin flips always involves a sign that 
can not be gauged away by a different choice of guiding wavefunctions}
   \label{fig3}
\end{figure}

For unfrustrated systems these loops do not exist and one can remove
the minus sign by a transformation of the spin operators
\begin{equation} \label{a6}
S^x_i \rightarrow -S^x_i, \quad \quad \quad
S^y_i \rightarrow -S^y_i, \quad \quad \quad
S^z_i \rightarrow S^z_i
\end{equation}
which leave the commutation operators invariant. Applying this transformation
on every other spin (the white fields of a 
checkerboard) all flips involving a pair of nearest neighbors then give a
positive matrix element for the Greens Function. So when $J_2=0$ the
appearant sign problem is transformed away. 
For sufficiently small $J_2$, Marshall \cite{Mar} has shown
that the wave function of the system has only positive components 
(after the ``Marshall'' sign flip (\ref{a6})).
So the minus sign problem is not due to the wave function but to the 
frustration. (For the Hubbard model it is the guiding wave function which
must have minus signs due to the Pauli principle, while the bare transition
probabilities can be taken positive).

Due to the minus sign the weight of a long path picks up a arbitrary sign.
Generally the weights are also growing along a path. Thus if various paths
are traced out by a number of independent walkers, the average over the
paths or the walkers becomes a sum over large terms of both
signs, or differently phrased: the average becomes small with respect to the
variance; the signal gets lost in the noise.

Ceperley and Alder \cite{Cep} constructed a method, Fixed Node Monte
Carlo (FNMC), which avoids the minus sign problem
at the expense of introducing an approximation. Their method is designed
for continuum systems and handling fermion wavefunctions. They argued that
the configuration space in which the wavefunction has a given sign, say
positive, is sufficient for exploring the properties of the groundstate,
since the other half of the configuration space contains identical information.
Thus they designed a method in which the walkers remain in one domain of
a given sign, essentially by forbidding to cross the nodes of the wavefunction.
The approximation is that one has to take the nodal structure of the guiding
wavefunction for granted and one cannot improve on that, at least not without
sophistocation (nodal release). The method is variational in the sense that
errors in the nodal structure always raise the groundstate energy.

It seems trivial to take over this idea to the lattice but it is not. The
reason is that in continuum systems one can make smaller steps when a walker
approaches a node without introducing errors. In a lattice system the
configuration space is discrete; so the location of the node is not
strictly defined. The important part is that, loosely speaking the nodes
are between configurations and one cannot make smaller moves than displacing
a particle over a lattice distance or flip a pair of spins. Van Bemmel et al.
\cite{Bem} adapted the FNMC concept to lattice systems preserving 
its variational character. This extension to the lattice suffers from the
same shortcoming as the method of Ceperley and Alder: the ``nodal''
structure of the guiding wavefunction is given and cannot be improved by the
Monte Carlo process. Recently Sorella \cite{Sor} proposed a modification which
overcomes this drawback. It is based on two ingredients.

Sorella noticed that the following effective hamiltonian yields also an upper
bound to the energy:
\begin{equation} \label{f1}
\begin{array}{rcl}
\langle R | \tilde{{\cal H}}_{\rm eff} | R' \rangle & = & 
\left\{ \begin{array}{rcl}
\langle R | \tilde{{\cal H}} | R' \rangle & {\rm if} & 
\langle R | \tilde{{\cal H}} | R' \rangle < 0 \\*[2mm]
-\gamma \langle R |  \tilde{{\cal H}} | R' \rangle & {\rm if} & 
\langle R | \tilde{{\cal H}} | R' \rangle >0  \quad (\gamma \geq 0)
\end{array} \right. \\*[6mm]
\langle R | \tilde{{\cal H}}_{\rm eff} | R \rangle & = & 
\langle R | \tilde{{\cal H}} | R \rangle  + (1 + \gamma) V_{\rm sf} (R)
\end{array}
\end{equation}
Here the ``sign flip'' potential is the same as that of ten Haaf et al.
\cite{Haa} and given by
\begin{equation} \label{e2}
V_{\rm sf} (R) = \sum_{R'_{\rm na}} \langle R'_{\rm na} | 
\tilde{{\cal H}} | R \rangle
\end{equation}
where the subscript ``na'' (not--allowed) on $R'$ restricts the summation 
to the moves for which the
matrix element of the hamiltonian is positive  (\ref{f1}). 

If the guiding
wavefunction were to coincide with the true wavefunction, the simulation
of the effective hamiltonian, which is sign free by construction, yields
exact averages. So one may expect that good guiding wavefunctions lead to
good upperbounds for the energy. This upperbound increases with $\gamma$,
indicating that $\gamma = 0$ seems the best choice, which is the effective 
hamiltonian of ten Haaf et al.\cite{Haa}. That hamiltonian however is a 
truncated version of the true hamiltonian in which all the dangerous moves 
are eliminated. The sign flip potential must correct this truncation by
suppressing the probability that the walker 
will stay in a configuration with a large potential. 

The second ingredient uses the fact that
the hamiltonian (\ref{f1}) explores a larger phase space and therefore
contains more information than the truncated one. Parallel to the
simulation of the effective hamiltonian one can calculate the weights for
the true hamiltonian. As we saw in the summary
of the GFMC method, forcefully made positive transition rates still
contain the correct weights when supplemented with sign functions.
For the true weights of the transition probabilities
as given in (\ref{f1}), the ``sign function'' must be chosen as
\begin{equation} \label{f2}
s(R,R') = \left\{ \begin{array}{lcl}
1 & {\rm if} & \langle R | \tilde{{\cal H}} | R' \rangle >0  \\*[4mm]
-1/\gamma & {\rm if} & \langle R | \tilde{{\cal H}} | R' \rangle < 0 \\*[4mm]
\displaystyle \frac{1 - \epsilon \langle R |\tilde{{\cal H}} |R \rangle}
{1 - \epsilon \langle R |\tilde{{\cal H}}_{\rm eff} |R \rangle} & 
{\rm if} & R = R'
\end{array} \right.
\end{equation}
With these ``signs'' in the weights a proper average can be calculated, but
these averages suffer from the sign problem, the more so the smaller $\gamma$
is as one sees from (\ref{f2}). So some intermediate value of $\gamma$
has to be chosen. Fortunately 
the results are not too sensitively dependent on $\gamma$; the value
$\gamma=0.5$ is a good compromise and has been taken in our simulations. 

In any simulation some walkers obtain a large weight and others a small one.
To lower the variance branching is regularly applied, which means a 
multiplication of the heavily weighted walkers in favor of the removal of
those with small weight. It is not difficult to do this in an unbiased way.
Sorella \cite{Sor} proposed to use the branching much more effectively in 
conjunction with the signs defined in (\ref{f2}). The average sign is an 
indicator of the usefulness of the set of walkers. Start with a set of 
walkers with positive sign. When the average sign becomes
unacceptably low, the process is stopped and a reconfiguration takes place.
The walkers are replaced by another set with positive weights only, such that
a number of measurable quantities gives the same average. The more observables
are included the more faithful is the replacement. The construction of the
equivalent set requires the solution of a set of linear equations. 
With the new set of walkers one continues the simulation on
the basis of the effective hamiltonian and one keeps track of the true
weights with signs. The reconfiguration on the basis of some observables
gives at the same time a measure for these obervables. Thus measurement
and reconfiguration go together. As the number
of observables that can be included is limited some biases are necessarily
introduced. Sorella showed that the error in the
energy of the guiding wave function is easily reduced  by a factor of 10,
whereas reduction by the FNMC of ten Haaf et al. \cite{Haa} rather gives
only a factor 2. 
\begin{figure}[h]   
   \centering \epsfxsize=\linewidth
   \epsffile{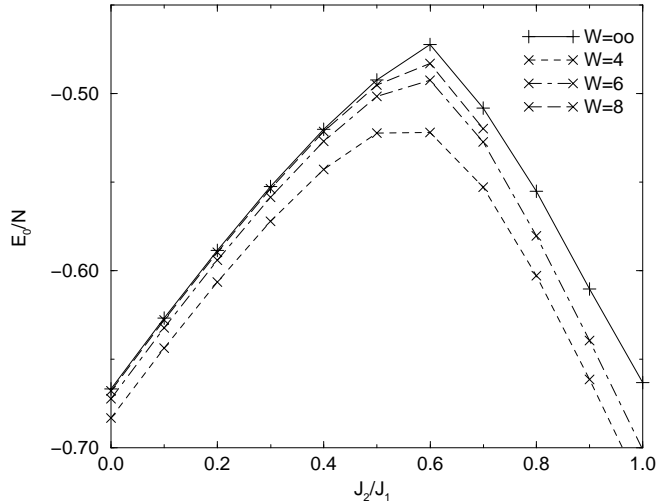}
   \caption{The energy as function of the frustration ratio}
   \label{fig4}
\end{figure}

\section{Results for the DMRG}

In this section we give a brief summary of the results of a pure 
DMRG--calculation. Extensive details can be found in \cite{Luc}.
The systems are strips of widths up to $W=8$ and of various lengths $L$.
They are periodic in the small direction and open in the long direction.
The periodicity enables us to study the spin stiffness. We have chosen 
open boundaries in the long direction to avoid the errors in the DMRG
wavefunction due to periodic boundaries. Since we have good control 
of the scaling behavior in $L$ we extrapolate to 
$L \rightarrow \infty$ \cite{Luc}. In the small direction
we are restricted to $W=2,4,6$ and 8 as odd values are not compatible
with the antiferromagnetic character of the system. For wider system sizes
the number of states which has to be taken into account exceeds the 
possiblities of the present workstations. Our criterion is that the value
of the energy does not drift anymore appreciably upon the inclusion of more
states. This does not mean that the wavefunction is virtually exact, 
since the energy is a rather insensitive probe for the wavefunction.
For instance correlation functions still improve from the
inclusion of more states.
In Fig. \ref{fig4} we present the energy as function of the ratio $J_2/J_1$,
for strip widths 4,6 and 8 together with the best extrapolation to
infinite width systems. 
The figure strongly suggests that the infinite system undergoes a
first order phase transition around a value 0.6.
This can be attributed to the transition to a columnar order (lines of 
opposite magnetisation). It is impossible to deduce more information 
from such an energy curve as other phase transitions are likely to be 
continuous with small differences in energy between the phases.

The spin stiffness can be calculated with the DMRG--wavefunction for
systems which are periodic in at least one direction \cite{Luc}.
\begin{figure}[h]   
   \centering \epsfxsize=\linewidth
   \epsffile{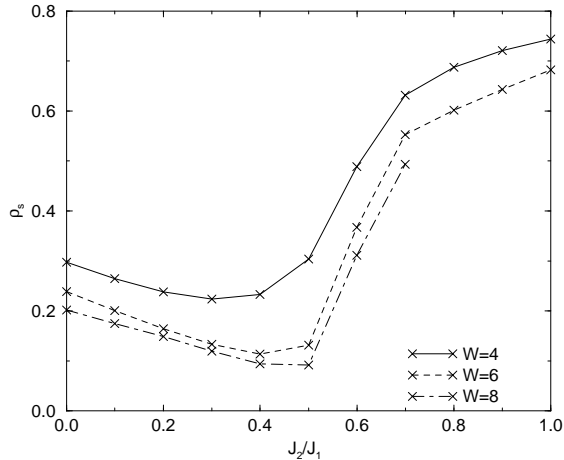}
   \caption{The stiffness ${\bf \rho}_s$ as function of the frustration ratio.
Finite size extrapolations put the region where $\rho_s$ vanishes between
0.38 and 0.62 \cite{Sch}}
   \label{Fig. 4 }
\end{figure}
The result of the computation is plotted in Fig. 5.
One observes a substantial decrease of $\rho_s$ in the frustrated
region indicating the appearance of a magnetically disordered phase. 
In contrast to the energy the data 
do not allow a meaningful extrapolation to large 
widths. The lack of clear finite size scaling behavior in the regime of 
small values of $W$ prevents to draw firm conclusions on the disappearence 
of the stiffness in the middle regime.

For the correlation functions following from the DMRG wavefunction we refer
to \cite{Luc}.

\section{Results for GFMC with SR}

We now come to the crux of this study: the simulations of the system with
GFMC, using the DMRG wavefunctions to guide the importance sampling.
All the simulations have been carried out for  $10 \times 10$ lattice 
with open boundaries. Standardly we have 6000 walkers and we run the 
simulations for about $10^4$ measurements. These measuring points are
not fully independent and the variance is determined by chopping up the
simulations into 50-100 groups, often carried out in parallel on different
computers. We first give an overall assessment of the correlation 
function pattern and then analyze some values of the ratio $J_2/J_1$. 

In the first series
we have used the guiding wavefunction on the basis of the meandering
path Fig. \ref{fig2}(b), because it gives a better energy than the 
straight option (a) . The number of basis states 
is $m=75$, which is small enough to carry out the
simulations with reasonable speed and large enough that trends begin
to manifest themselves. Measurements of a number of correlation functions are 
made in conjunction with Stochastic Reconfiguration as described in section 7.
The details of these calculations are given in Table \ref{tab0}.
Note that the DMRG guiding wavefunction gives a better energy for
the meandering path than for the straight path for values of
$J_2/J_1$ up to 0.6. From 0.7 on this difference is virtually absent.
This undoubtly has to do with the change to the columnar state which
can equally well be realized by both paths. The value of $\epsilon$ has
been chosen as a compromise: independent measurements require a large
$\epsilon$ but the minus sign problem requires to apply often Stochastic 
Reconfiguration i.e. a small $\epsilon$. One sees that in the heavily 
frustrated region the $\epsilon$ must be taken small. In fact the more
detailed calculations for $J_2 = 0.3J_1$ and $J_2 = 0.5J_1$ were carried 
out with $\epsilon = 0.01$.

In Fig. 6 and 7 we have plotted a sequence of visualizations of the 
correlations. From top to bottom (zig--zag)
they give the correlations for the values of $J_2/J_1$. In order
to highlight the differences a distinction is made between correlations
which are above average (solid lines) and below average (broken lines).
All nearest neighbor spin correlations shown are negative. 
In all the pictures one sees the influence of the boundaries on the spin
correlations. Only 1/4 of the lattice has been pictured, the other
segments follow by symmetry. The upper right corner, which
corresponds to the center of the lattice, is the most significant for the
behavior of the bulk. The overall trend is that spatial variations
in the correlation functions occur in growing size with $J_2/J_1$.
On the side of low $J_1/J_2$ (N\'eel phase)
one sees dimer patterns in the horizontal direction, they
turn over to vertical dimers (around $J_2 = 0.7 J_1$) and rapidly
disappear in the columnar phase.
This is again support for the fact that the columnar phase is separated
from the intermediate state by a first order phase transition.

Open boundary conditions have the disadvantage of boundary effects, which
make it more difficult to distinguish between spontaneous and induced 
breaking of the translational symmetry. On the other hand for open 
boundaries, dimers, plaquettes or any other interruption of the 
translational symmetry have a natural reference frame. 
The correlations are not only influenced by the boundaries of the system,
also the guiding DMRG--wavefunction leaves its imprint on the results.
This is mainly due to the fact that we have only mixed estimators for the
correlation functions, which show a mix of the guiding wavefunction and the
true wavefunction. The improved estimator, used in these pictures, corrects
for this effect to linear order in the deviation.\footnote{Forward walking
allows to make a pure estimate of the correlations, but requires much
more calculations. \cite{Sor2}} The ladder like structure
in the DMRG path is reflected in a ladder like pattern in the correlations
as an inspection of the correlations in the DMRG wavefunctions (not shown
here) reveals. But ladders are clearly also present in the GFMC results 
shown in the pictures. 

In order to eliminate the influence of the guiding 
wavefunction we scrutinize some of values of $J_2/J_1$ in more detail,
by inspecting how the results depend on the size of the basis in the
DMRG wavefunction and on the choice of the DMRG path. Since we are
mostly interested in the behavior of the infinite lattice, we discuss mainly
the behavior of the correlations in and around the central plaquette. 
So we study a sequence of DMRG wavefunctions for $m=32$, 75, 100, 128 
and 150(200) and carry out for each of them extensive GFMC simulations.
First we look to the case $J_2=0$, which is easy because we know 
that it must be N\'eel ordered and therefore it serves as a 
check on the calculations. Then we take $J_2=0.3 J_1$ which is the
most difficult case since it is likely to be close to a phase transition.
Finally we inspect $J_2 = 0.5 J_1$ where  
we are fairly sure that some dimerlike phase is realized. 

\subsection{$J_2=0$}

For the unfrustrated Heisenberg model we have several checkpoints for
our calculations. We can find to a high degree of accuracy the groundstate
energy and we are sure that the N\'eel phase is homogeneous, i.e. that 
the correlations show no spatial variation other than that of the 
antiferromagnet.
We have two ways of estimating the energy of a $10 \times 10$ lattice.
The first method is based on finite size interpolation. From 
DMRG calculations \cite{Luc} we have an exact value for a
$4 \times 4$ lattice, an accurate value for the $6 \times 6$ lattice and
a good value for the $8 \times 8$ lattice. There is also the very  
accurate calculation of Sandvik \cite{San} for an infinitely large lattice,
yielding the value of $e_0=-0.669437(5)$. 
The leading finite size correction goes as $1/L$. 
Including also a $1/L^2$ term we have esimated the value for a $10 \times 10$
lattice as 0.629(1) and incorporated this value in Table \ref{tab1}(a). 
We stress that this is an {\it interpolation} for which the value of Sandvik 
is the most important input. 

The second method is less well founded and uses the experience that 
DMRG energy estimates can be improved considerably by {\it extrapolating} 
to zero truncation error. When plotted as function of this
truncation error the energy is often remarkably linear. In Table 
\ref{tab1}(b) we give for a series of bases $m=32, 75, 100, 128$
and 150, the values of the truncation error and the corresponding DMRG energy
per site together with the extrapolation on the basis of linear behavior.\footnote{The value for $m=150$ is not in line with the others. This can be
explained by the fact that the construction of this DMRG wavefunction was
slightly different from the others in which the basis was built up gradually.}
Note that the two estimates are compatible. In Table \ref{tab1}(b) we have 
also listed the values of the GFMC simulations for the corresponding values 
of $m$. They do agree quite well with these estimates in particular with the 
one based on finite size scaling. We point out that one 
would have to go very far in the number of states in the DMRG calculation to
obtain an accuracy that is easily obtained with GFMC. Thus the combination
of GFMC and DMRG does really better than the individual components.
One might wonder why there is still a drift to lower energy values in
the GFMC simulations (which is also present in the tables to come).
The reason is that the DMRG wavefunction is strictly zero outside a 
certain domain of configurations, because the truncation of the basis
involves also the elimination of certain combinations of conserved 
quantities of the constituing parts. The domain of the wavefunction grows
with the size of the basis.

Turning now to the correlations it seems that they are homogeneous in 
the center of the lattice for $J_2=0$. However a closer inspection
reveals small differences. In Table \ref{tab2} we list the asymmetries in
the horizontal and vertical directions of the spin correlations in and around
the central plaquette as function of the number of states.
If we number the spins on the lattice as 
${\bf S}_{n,m}$ with $1 \leq n,m \leq 10$, the central plaquette 
has the coordinates (5,5), (5,6), (6,5) and (6,6). We then define the 
asymmetry parameters $\Delta_x$ and $\Delta_y$ as
\end{multicols}
\begin{equation} \label{g1}
\left\{ \begin{array}{l}
\Delta_x  =  
\frac{1}{4} \langle {\bf S}_{4,5} \cdot {\bf S}_{5,5} +
{\bf S}_{4,6} \cdot {\bf S}_{5,6} + {\bf S}_{6,5} \cdot {\bf S}_{7,5} +
{\bf S}_{6,6} \cdot {\bf S}_{7,6} \rangle -\frac{1}{2} \langle
{\bf S}_{5,5} \cdot {\bf S}_{6,5} + {\bf S}_{5,6} \cdot {\bf S}_{6,6}
\rangle \\*[2mm]
\Delta_y  = 
\frac{1}{4} \langle {\bf S}_{5,4} \cdot {\bf S}_{5,5} +
{\bf S}_{6,4} \cdot {\bf S}_{6,5} + {\bf S}_{5,6} \cdot {\bf S}_{5,7} +
{\bf S}_{6,6} \cdot {\bf S}_{6,7} \rangle -\frac{1}{2} \langle
{\bf S}_{5,5} \cdot {\bf S}_{5,6} + {\bf S}_{6,5} \cdot {\bf S}_{6,6} \rangle
\end{array} \right.
\end{equation}
\begin{multicols}{2}
So $\Delta_x$ is the average value of the 
correlations on the 4 horizontal bonds which are connected to the central
plaquette minus the average of the values on the 2 horizontal bonds in
the plaquette. Similarly $\Delta_y$ corresponds to the vertical direction.
The values for the asymmetry in Table \ref{tab2} in the vertical direction 
are so small that they have no significance. Note that the
anticipated decrease in $\Delta_x$ is slow in DMRG and therefore also
slow in the mixed estimator of the GFMC. The improved estimator (\ref{d15})
however is truely an improvement! So one sees that all the observed small
deviations from the homogeneous state will disappear with the increase of 
the number of states in the basis of the DMRG wavefunction. (In general the
accuracy of the correlations is determined by that of the GFMC simulations.
We get as variance a number of the order 0.01, implying twice that value
for the improved estimator) $\;$ The vanishing of $\Delta_x$ and $\Delta_y$
also prove that finite size effects are small in the center of the 
$10 \times 10$ lattice. 
From these data we may conclude that the GFMC can make up for
the errors in the DMRG wavefunction for a relative low number of basis states.
We have not carried out a similar series for the straight path since
this will certainly show no dimers as will become clear from the following
cases.

\subsection{$J_2 = 0.3 J_1$}

This case is the most difficult to analyze since it is expected to be close
to a continuous phase transition from the N\'eel state to a 
dimerlike state. As is known \cite{Ost} the DMRG structure of the wavefunction
is not very adequate to cope with the long--range correlation in the
spins typical for a critical point. In Table \ref{tab3} we have presented the
same data as in Table \ref{tab2} but now for $J_2=0.3$.
There is no pattern in the energy as function of the truncation error
$\delta$. The decrease of the energy as function of the size of the 
basis $m$ is in the DMRG wavefunctions is not saturated.
The GFMC simulations lead to a notably lower energy and they
do hardly show a leveling off as function of the basis of the 
guiding wavefunction. All these points
are indicators that the DMRG wavefunction is rather far from convergence
and that more accurate data would require a much larger basis. As far as the 
staggering in the correlations is concerned the values for $\Delta_x$ are
significant, also because the simulation results generally increase the 
values. Those for $\Delta_y$ are not small enough to be considered as noise.
Given the fact that most authors locate the
phase transition at higher values $J_2 \simeq 0.4 J_1$ we would expect
both $\Delta$'s to vanish. So either the dimerlike state is realized for
values as low as $J_2 = 0.3 J_1$ or dimer formation already starts 
in the N\'eel state. 

To get more insight in the nature of the groundstate we have also carried
out the same set of simulations on the straight path (a) in Fig. \ref{fig2}. 
This guiding wavefunction shows virtually no formation of dimers in 
any direction as can be observed from Table \ref{tab4}.
In spite of the fact that the trends indicated in the table have not come to
convergence one may draw a few conclusions from the comparison of the
two sets of simulations. The overal impression is that the meandering
guiding wavefunction represents a groundstate of a different symmetry
as compared to the straight path guiding wavefunction.
The meandering wavefunction prefers dimers in the horizontal direction
and the straight wavefunction leads to some dimerization in the vertical
direction. The difference  also shows up in the energy, it is
not only large on the DMRG level but it also persists at the GFMC level. 
We see similar trends in the next case.

\subsection{$J_2 = 0.5 J_1$}

By any estimate this value of the next nearest neighbor coupling leads to
a dimerlike state if it exists at all. No accurate data are available 
on the energy of the $10 \times 10$ system to compare to our results.
In Table \ref{tab5} we list the data for a set of DMRG wavefunctions 
with bases $m=32$, 75, 100, 128, 150 and 200.
The DMRG values of the energy (with exception of the value for $m=32$) can
be extrapolated to zero truncation error with the limiting value
$E_0 = -48.4(1)$, which corresponds very well with the level in the
GFMC values for larger sizes of the basis.
This indicates again that the GFMC simulations can
make up for the shortcoming of the DMRG wavefunction. One would indeed have
to enlarge the basis to $m$ of the order of 1000 in order to achieve the 
value of the energy of the simulations which use DMRG guiding 
wavefunctions with a basis of the order of 100. 

The staggering in the correlations expressed by the quantities $\Delta_x$ 
for the horizontal direction and $\Delta_y $ for the vertical direction,
has values that are significant. If one looks to the contributions 
of the DMRG wavefunction and the GFMC simulation separately, one observes
that the overall values do agree quite well, with the tendency that the
GFMC simulations lowers the staggerring in the horizontal direction and
slightly increases it in the vertical direction.
So we may conclude that indeed in the groundstate of the $J_2=0.5 J_1$ system, 
the correlations of the spins are not translation invariant but show 
a staggering. However these results neither confirm the picture that
the dimerstate is the lowest (as suggested by Kotov et al. \cite{Kot})
nor that the plaquettestate is the groundstate (as concluded by Capriotti 
and Sorella \cite{Sor2}). We comment on these discrepancies further in the
discussion.

Again it is worthwhile to compare these results with a simulation on the
basis of the straight path (a) in Fig. \ref{fig2}. 
Here it is manifest that the straight path prefers to have the
dimers in the vertical direction. Again the impression is that the
straight path leads to a different symmetry as compared to the meandering
path. It is not only the different preference in the main direction of 
the dimers, also the secondary dimerization in the perpendicular direction,
notably in the meandering case, is not present in the straight case.
The fairly large difference in energy on the DMRG level becomes quite small
on the GFMC level.
 
\section{Discussion}

We have presented a method to employ the DMRG wavefunctions as guiding
wavefunctions for a GFMC simulation of the groundstate. Generally 
the combination is much better than the two individual methods. The
GFMC simulations considerably improve the DMRG wavefunction. In the
intermediate regime the properties of the GFMC simulations depend 
on the guiding wavefunction as the results for two different DMRG guiding
wavefunctions show. 

\begin{figure}[h]
  \centering \epsfig{file=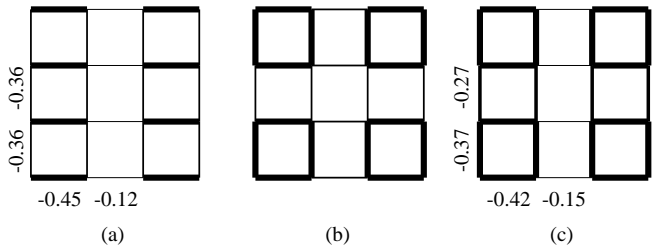,width=\linewidth}
 \caption[]{The correlation pattern for the nearest spins for $J_2 = 0.5J_1$;
(a) according to Kotov et al. \cite{Kot2}: a dimer pattern in which the 
strength of the correlation is indicated; (b) according to Capriotti and 
Sorella \cite{Sor2}: a plaquette state and (c) according to this paper: an
intermediate pattern in which the translational invariance is broken in
both directions but with unequal strength. The values indicated are those
based on the meandering path and the improved estimator.}
\label{comp}
\end{figure}

The method has been used to observe spin correlations in the frustrated
Heisenberg model on a square lattice. In this discussion we focus 
on the intermediate region where the model is most frustrated
and which is the ``piece de resistance'' of the present research.
We see patterns of strongly correlated nearest neighbor spins, to be
called dimers. To indicate what me mean by strong and weak we give the
values in and around the central square of the $10 \times10$ lattice, 
for the case $J_2 = 0.5 J_1$. In Fig. \ref{comp}(c) we have given the
values of the central square extrapolated to an infinite lattice.
The values are based on the improved estimator and it is interesting to
see the trends. The horizontal strong correlation of -0.42 is the result
of the DMRG value -0.44 and the GFMC value -0.43, while the weak bond
-0.15 is the result of the DMRG value -0.09 and the GFMC value -0.12. 
Thus the GFMC weakens the order parameter $D_x$ associated with the 
staggering. For the vertical direction
there is hardly a change from DMRG to GFMC. One has to go to the next 
decimal to see the difference. The strong bond equals -0.368 and is 
coming from the DMRG value -0.375 and the GFMC
value -0.371, while the improved weak bond of -0.271 is the resulting value
of -0.275 for DMRG and -0.273 for GFMC. 

Before we comment on this result we discuss the influence of the choice
of the guiding wave function. We note that for both points $J_2 = 0.3 J_1 $
and $J_2 = 0.5 J_1 $ the two choices for the DMRG wavefunction give
different results. First of all the main staggering is for the meandering
path (b) of Fig. \ref{fig2} in the horizontal direction, while the 
straight path (a) of Fig. \ref{fig2} prefers the dimers in the vertical 
direction. There is
not much difference in the values of the strong and weak correlations.
Secondly the straight path shows no appreciable staggering in the other
direction, so one may wonder whether the observed effect for the meandering
path is real. In our opinion this difference has to do with the effect that 
the DMRG wavefunction ``locks in'' on a certain symmetry. The straight path
yields a groundstate which is truely dimerlike in the sense that it is
translational invariant in the direction perpendicular to the dimers.
The meandering path locks in on a different groundstate which holds the
middle between a dimerlike and a plaquettelike state. The GFMC simulations
cannot overcome this difference in symmetry, likely because the two
lowest states with different symmetry are virtually orthogonal. On the
DMRG level there is a large difference in energy between the two states,
favoring the meandering path strongly, on the GFMC level this difference
has become very small. With this observation in mind we compare
our result with other findings.

The results of the series expansions \cite{Kot}, \cite{Kot2} and 
\cite{Sin} are shown in Fig. \ref{comp}(a).
Their correlations organize themselves in spinladders.
The correlations on the rungs of the ladder are $-0.45 \pm 0.5$ which compares
well with our strongest horizontal correlation and this holds also for
the weak horizontal correlation (--0.12 vs --0.15). The most noticeble
difference is the value of our weak correlation in the vertical direction
(--0.27 vs --0.36) while the strong correlation (--0.37 vs --0.36) agrees.
There is no real conflict between our result and theirs since 
the symmetry they find is fixed by the state around which the series
expansion is made. So our claim is only that our state with different 
symmetry is the lower one. In fact in the paper of Singh et al. \cite{Sin},
it is noted that the susceptibility to a staggering operator in the 
perpendicular direction (our $\Delta_y$) becomes very large in the dimer 
state for $J_2 = 0.5 J_1$ which we take as an indication of the nearby 
lower state. The analytical calculations in \cite{Kot} and \cite{Kot2}
however do not support the existence of the state we find.

Neither do we find support for the plaquette state found in \cite{Sor2},
which we have sketched in Fig. \ref{comp}(b).
The evidence of this investigation is based on the boundedness of the
susceptibility for the operator which breaks the orientational symmetry and
the divergence of the susceptibility for the order parameter breaking
translational invariance (corresponding to $\Delta_x$). They have not
separately investigated the values of $\Delta_x$ and $\Delta_y$ since
their groundstate has the symmetry of the lattice and one would 
find automatically the same answer.
They conclude that in absence of orientational order parameter and with the
presence of the translational order parameter the state must be plaquettelike.
We believe that their result is influenced by the guiding wavefunction
for which the one-step Lanczos approximation is taken. This wavefunction
certainly has the symmetry of the square and again GFMC cannot find
a groundstate with a different symmetry. 

Finally we comment on the fact that we find the dimerization already 
for values as low as $J_2 = 0.3 J_1$ at least for the meandering path.
As we have mentioned earlier the results as function of the number of
states have not sufficiently converged to make a firm conclusion, the
more so since there is a large difference between DMRG and GFMC. Still
it could be an indication that the phase transition from the N\'eel
state to the dimer state takes place for lower values than the
estimated $J_2 = 0.38 J_1$ \cite{Sch}.
 
Thus many questions are left over, amongst others how the order parameters
behave as function of the frustation ratio in the intermediate region. We
feel that the combination of DMRG and GFMC is a good tool to investigate
these issues since they demonstrate {\it ad oculos} the correlations in the
intermediate state.

{\bf Acknowledgement} 
The authors are indebted to Steve White for making his software available.
One of us (M. S. L. duC. de J.) gratefully acknowledges the hospitality
of Steve for a stay at Irvine of 3 months, where the basis 
of this work was laid. The authors have also benefitted from 
illuminating discussions with Subir Sachdev and Jan Zaanen. The authors
want to acknowledge the efficient help of Michael Patra with the 
simulations on the cluster of PC's of the Instituut-Lorentz.

\end{multicols}
\vspace*{1cm}

\begin{table}[h]
  \begin{center}
\begin{tabular}{|l|l|l|l|l|l|}
 \hline
 & & \multicolumn{2}{|c|}{} & \multicolumn{2}{|c|}{} \\*[-3mm]
 & & \multicolumn{2}{|c|}{Straight} & \multicolumn{2}{|c|}{Meander} \\*[1mm]
\cline{2-6}
$J_2$ & $\epsilon$& $E_{\rm DMRG}$ & $E_{\rm GFMC}$ & $E_{\rm DMRG}$ & $E_{\rm GFMC}$ \\*[1mm]
\hline
0.0 & 0.3    & -61.30 & -62.33(8)  & -61.84 & -62.54(4) \\*[1mm]
0.1 & 0.06   & -57.96 &            & -58.53 & -59.25(2) \\*[1mm]
0.2 & 0.04   & -54.75 & -56.08(11) & -55.48 & -56.22(4) \\*[1mm]
0.3 & 0.02   & -51.75 & -53.17(4)  & -52.50 & -53.38(3) \\*[1mm]
0.4 & 0.02   & -49.00 & -50.51(8)  & -49.92 & -50.60(5) \\*[1mm]
0.5 & 0.014  & -46.68 & -47.76(6)  & -47.78 & -48.34(4) \\*[1mm]
0.6 & 0.015  & -45.41 &            & -46.03 & -46.40(3) \\*[1mm]
0.7 & 0.015  & -45.67 &            & -45.60 & -46.00(2) \\*[1mm]
0.8 & 0.02   & -49.16 &            & -49.13 & -49.60(9) \\*[1mm]
0.9 & 0.02   & -53.61 &            & -53.70 & -54.52(2) \\*[1mm]
1.0 & 0.02   & -58.46 & -59.71(9)  & -58.64 & -59.80(8)\\
\hline
\end{tabular}
\end{center}
\caption{\label{tab0} For each degree of frustration the imaginary 
time interval $\epsilon$, the energy of the guiding state 
$E_{\rm DMRG}$ and that of the GFMC state $E_{\rm GFMC}$ are listed.}
\end{table}

\begin{table}[h]
\begin{center}
\begin{tabular}{|l|l|}
\hline
 &  \\*[-3mm]
$L$ & $e_0(L \times L)$ \\*[1mm]
\hline
 &  \\*[-3mm]
4 & -0.5740     \\*[1mm]
6 & -0.6031    \\*[1mm]
8 & -0.6188    \\*[1mm]
10 & -0.629(1)  \\*[1mm]
$\infty$ & -0.669437(5)    \\*[1mm]
\hline
\end{tabular} \quad \quad \quad
\begin{tabular}{|l|l|l|l|}
\hline
 & & &  \\*[-3mm]
 \# states & trunc. error & $ e_0$ (DMRG) & $ e_0$ (GFMC)\\*[1mm]
\hline
 & & &  \\*[-3mm]
32 &  21.2 $\times 10^{-5}$ & -0.6084 & -0.6192(1) \\*[1mm]
75 & 12.0 $\times 10^{-5}$  & -0.6184 & -0.6254(5) \\*[1mm]
100 & 10.5 $\times 10^{-5}$ & -0.6201 & -0.625(2)  \\*[1mm]
128 & 8.7 $\times 10^{-5}$  & -0.6214 & -0.6269(6)  \\*[1mm]
150 & 9.6 $\times 10^{-5}$  & -0.6231 & -0.6277(5)  \\*[1mm]
$2^N$ & 0  &  -0.631(3)       &      \\*[1mm]
\hline
\end{tabular}\\*[4mm]
\hspace*{-2cm} (a) \hspace*{7cm} (b)
\caption{\label{tab1} Interpolation (a) and extrapolation (b) 
estimates of the energy per site of a $10 \times 10$ lattice}
\end{center}
\end{table}
\vspace*{1cm}

\begin{table}[h]
\begin{center}
\begin{tabular}{|c|c|c|c|r|r|r|}
\hline 
  & \multicolumn{3}{|c|}{} & \multicolumn{3}{|c|}{} \\*[-3mm]
\# states  & \multicolumn{3}{|c|}{$\Delta_x$} & 
\multicolumn{3}{|c|}{$\Delta_y$} \\*[1mm]
\cline{2-7} 
 & & & & & & \\*[-3mm]
m & DMRG & GFMC & Improved & DMRG & GFMC & Improved \\*[1mm]
32 & 0.14373 & 0.09981 & 0.05589 & -0.00060 & 0.00078 & 0.00216 \\*[1mm]
75 & 0.07291 & 0.05668 & 0.04045 & 0.00081 & 0.00601 & 0.01121 \\*[1mm]
100 & 0.06432 & 0.04255 & 0.03088 & 0.00030 & 0.00173 & 0.00316 \\*[1mm]
128 & 0.05619 & 0.03734 & 0.01849 & 0.00091 & -0.00040 & -0.00173  \\*[1mm]
150 & 0.05044 & 0.03612 & 0.02221 & 0.00079 & 0.00261  & 0.00442  \\
\hline 
\end{tabular}\\*[4mm]
\caption{\label{tab2} Values for the asymmetry in the center for $J_2 = 0$.
As discussed in the text the error in the improved estimator values is of
the order 0.02, which means that for $m=128$ and higher the values are
statistically indistinguishable from zero.}
\end{center}
\end{table}
\vspace*{1cm}

\begin{table}[h]
\begin{center}
\begin{tabular}{|c|c|c|c|c|c|c|c|}
\hline
 &  \multicolumn{4}{|c|}{} & \multicolumn{3}{|c|}{} \\*[-3mm]
\# states  & \multicolumn{4}{|c|}{{\rm DMRG}} & 
\multicolumn{3}{|c|}{{\rm GFMC}} \\*[2mm]
\cline{2-8} 
 & & & & & & & \\*[-3mm]
$m$ & $\delta * 10^5$ & $E_{{\rm DMRG}} $ & $\Delta_x $ & 
$ \Delta_y $ & $E_{{\rm GFMC}} $ &  $\Delta_x $ & $ \Delta_y $ \\*[2mm]
\hline
 & & & & & & &\\*[-3mm]
32 & 19.0 & -51.609 & 0.27784 & 0.00295 & -52.81(43) & 0.363 & -0.009 \\*[2mm]
75 & 10.6 & -52.581 & 0.15462 & 0.00616 & -53.29(05) & 0.207 & 0.011 \\*[2mm]
100 & 9.4 & -52.707 & 0.14709 & 0.00943 & -53.32(33) & 0.145 & 0.009 \\*[2mm]
128 & 10.6 &-52.821 & 0.13042 & 0.00577 & -54.01(04) & 0.254 & 0.063 \\*[2mm]
150 & 10.4 &-52.888 & 0.12564 & 0.00737 & -54.10(12) & 0.236 & 0.103 \\*[1mm] 
\hline
\end{tabular} \\*[4mm]
\caption{\label{tab3} Energies and asymmetries for the case 
$J_2 = 0.3 J_1$ as function of the number of basis states $m$. $\delta$ is the
truncation error. The asymmetries $\Delta_x $ and $\Delta_y$ for the
GFMC simulations are 
calculated with the improved estimator. The guiding wavefunction is obtained
from the meandering path (b) in Fig. \ref{fig2}. The statistical error in
$\Delta_x$ and $\Delta_y$ is of the order 0.02}
\end{center}
\end{table} 

\begin{table}[h]
\begin{center}
\begin{tabular}{|c|c|c|c|c|c|c|c|}
\hline
 &  \multicolumn{4}{|c|}{} & \multicolumn{3}{|c|}{} \\*[-3mm]
\# states  & \multicolumn{4}{|c|}{{\rm DMRG}} & \multicolumn{3}{|c|}{{\rm GFMC}} \\*[2mm]
\cline{2-8} 
 & & & & & & & \\*[-3mm]
$m$ & $\delta * 10^5$ & $E_{{\rm DMRG}} $ & $\Delta_x $ & 
$ \Delta_y $ & $E_{{\rm GFMC}} $ &  $\Delta_x $ & $ \Delta_y $ \\*[2mm]
\hline
 & & & & & & &\\*[-3mm]
32 & 30.0 & -50.672 & 0.00032 & 0.01657 & -52.15(11) & 0.061 & 0.047 \\*[2mm]
75 & 18.9 & -51.733 & -0.00295 & 0.00426 & -53.21(10) & -0.030 & 0.036 \\*[2mm]
100 & 19.9 & -52.066 & 0.00349 & 0.00492 & -53.84(72) & 0.061 & 0.079 \\*[2mm]
128 & 24.6 & -52.302 & 0.00139 & 0.00791 & -53.50(19)& 0.079 & 0.027  \\*[2mm]
150 & 25.7 & -52.455 & 0.00222 & 0.00780 & -53.52(10) & 0.022  & 0.065  \\
\hline 
\end{tabular} \\*[4mm]
\caption{\label{tab4} Comparison of the energies and the 
values for the asymmetry in the center for the DMRG wavefunction based
on the first (straight) path (a) in Fig. \ref{fig2} and the associated 
GFMC simulation; $J_2 = 0.3 J_1$}
\end{center}
\end{table}
\vspace*{1cm}
\begin{table}[h]
\begin{center}
\begin{tabular}{|c|c|c|c|c|c|c|c|}
\hline
 &  \multicolumn{4}{|c|}{} & \multicolumn{3}{|c|}{} \\*[-3mm]
\# states  & \multicolumn{4}{|c|}{{\rm DMRG}} & \multicolumn{3}{|c|}{{\rm GFMC}} \\*[2mm]
\cline{2-8} 
 & & & & & & & \\*[-3mm]
$m$ & $\delta * 10^5$ & $E_{{\rm DMRG}} $ & $\Delta_x $ & 
$ \Delta_y $ & $E_{{\rm GFMC}} $ &  $\Delta_x $ & $ \Delta_y $ \\*[2mm]
\hline
 & & & & & & &\\*[-3mm]
32 & 11.8 & -47.116 & 0.43245 & 0.14667 & -47.55(29) & 0.295 & 0.065 \\*[2mm]
75 & 17.4 & -47.771 & 0.38954 & 0.13059 & -48.22(04) & 0.339 & 0.070 \\*[2mm]
100 &12.4 & -47.924 & 0.39364 & 0.07877 & -48.37(22) & 0.310 & 0.110 \\*[2mm]
128 & 8.4 & -48.014 & 0.37317 & 0.08246 & -48.32(05) & 0.336 & 0.139 \\*[2mm]
150 & 8.3 & -48.088 & 0.35819 & 0.07983 & -48.33(12) & 0.324 & 0.112 \\*[2mm]
200 & 7.6 & -48.153 & 0.34590 & 0.09973 & -48.43(05) & 0.272 & 0.094 \\*[1mm]
\hline
\end{tabular} \\*[4mm]
\caption{\label{tab5} Energies and asymmetries for $J_2=0.5 J_1$ with 
guiding wavefunction based on the meandering path (b) in Fig. \ref{fig2}}
\end{center}
\end{table}
\vspace*{1cm}

\begin{table}[h]
\begin{center}
\begin{tabular}{|c|c|c|c|c|c|c|c|}
\hline
 &  \multicolumn{4}{|c|}{} & \multicolumn{3}{|c|}{} \\*[-3mm]
\# states  & \multicolumn{4}{|c|}{{\rm DMRG}} & \multicolumn{3}{|c|}{{\rm GFMC}} \\*[2mm]
\cline{2-8} 
 & & & & & & & \\*[-3mm]
$m$ & $\delta * 10^5$ & $E_{{\rm DMRG}} $ & $\Delta_x $ & 
$ \Delta_y $ & $E_{{\rm GFMC}} $ &  $\Delta_x $ & $ \Delta_y $ \\*[2mm]
\hline
 & & & & & & &\\*[-3mm]
32 & 69.4 & -45.756 & 0.00172 & 0.24701 & -47.45(08) & 0.074 & 0.185 \\*[2mm]
75 & 26.2 & -46.718 & 0.00171 & 0.34950 & -47.81(25) & -0.025 & 0.302 \\*[2mm]
100 & 21.2 & -46.993 & 0.00063 & 0.33131 & -48.16(06) & -0.003 & 0.350 \\*[2mm]
128 & 24.6 & -47.231 & -0.00029 & 0.32994 & -48.31(08)& 0.013 & 0.291  \\*[2mm]
150 & 25.7 & -47.379 & 0.00215 & 0.32458 & -48.33(06) & -0.026  & 0.257  \\
\hline 
\end{tabular} \\*[4mm]
\caption{\label{tab6} Same as Table \ref{tab5} but
now for the ``straight'' path Fig (a) \ref{fig2}}
\end{center}
\end{table}

\begin{figure}  
  \epsfig{file=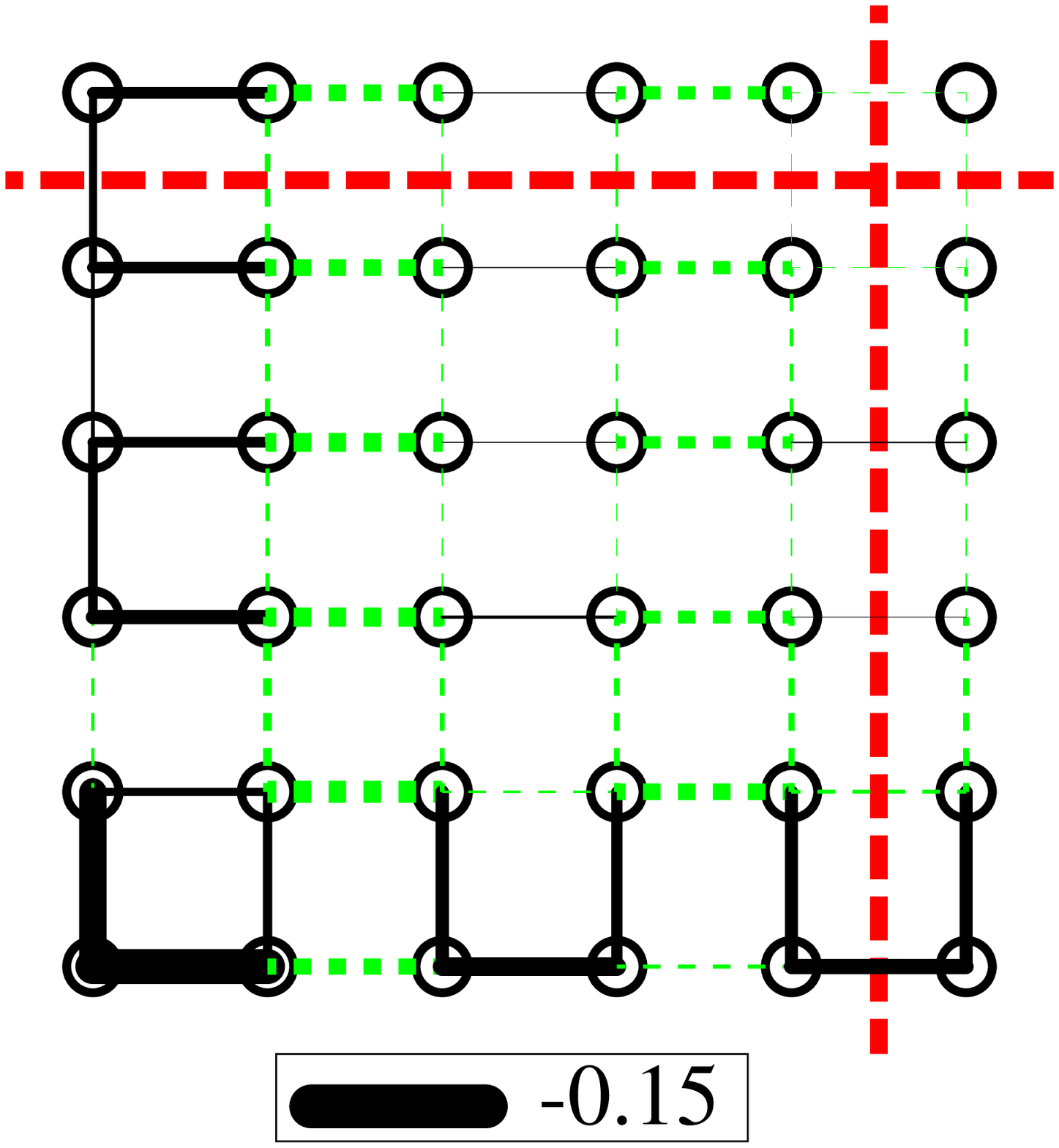,width=6cm}
  \hfill
  \epsfig{file=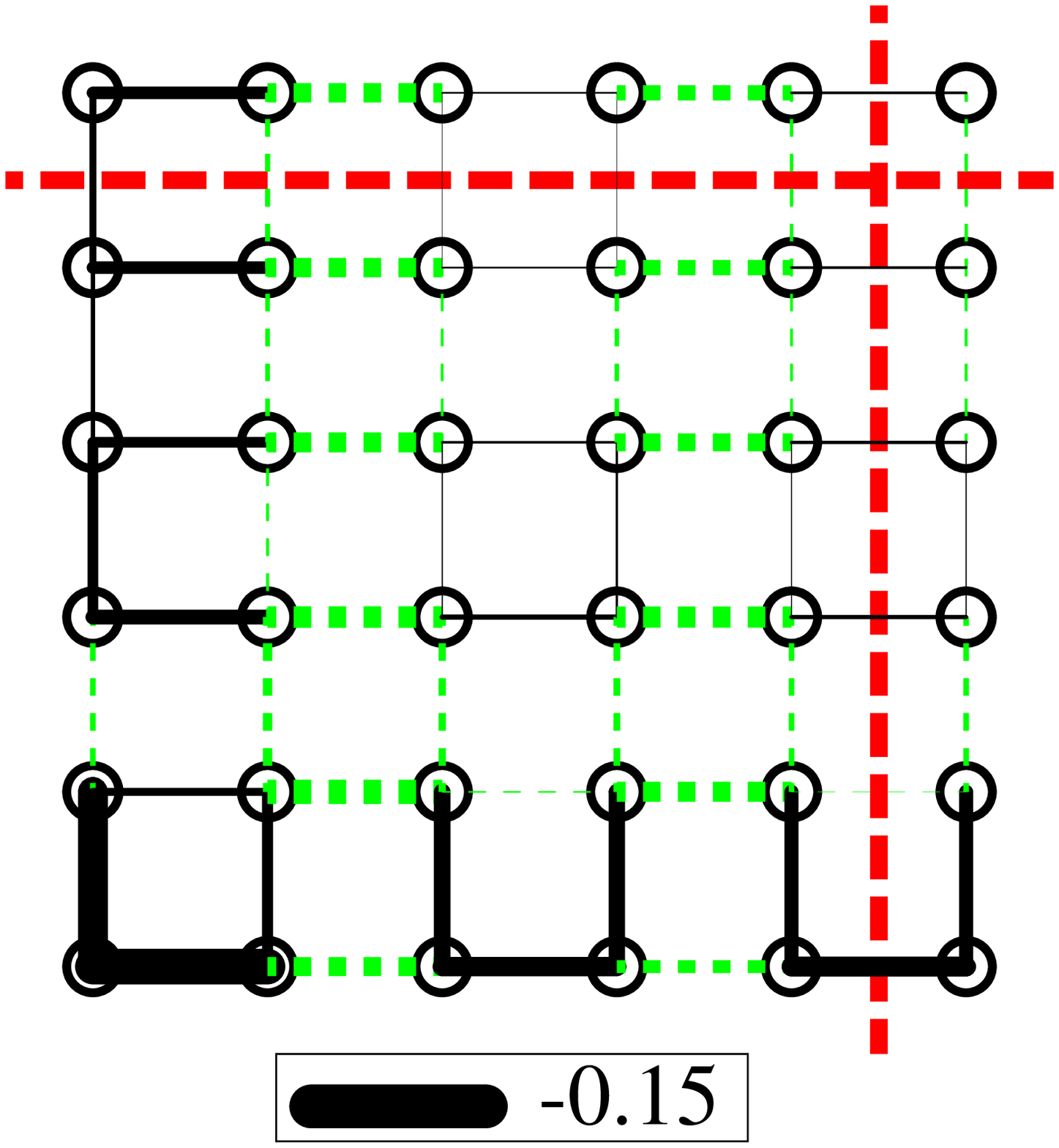,width=6cm} \vspace{0.5cm}\\
  \epsfig{file=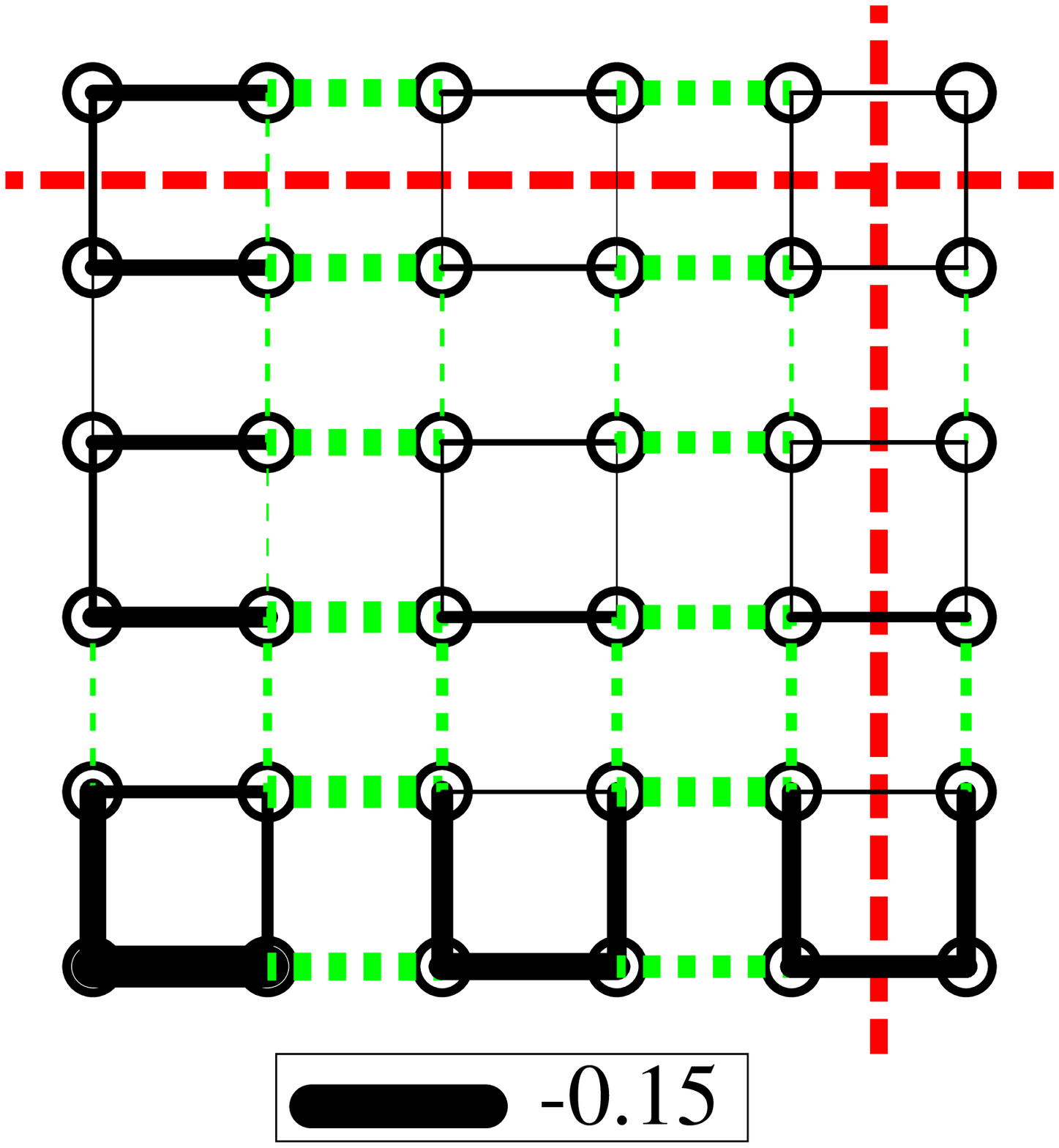,width=6cm}
  \hfill
  \epsfig{file=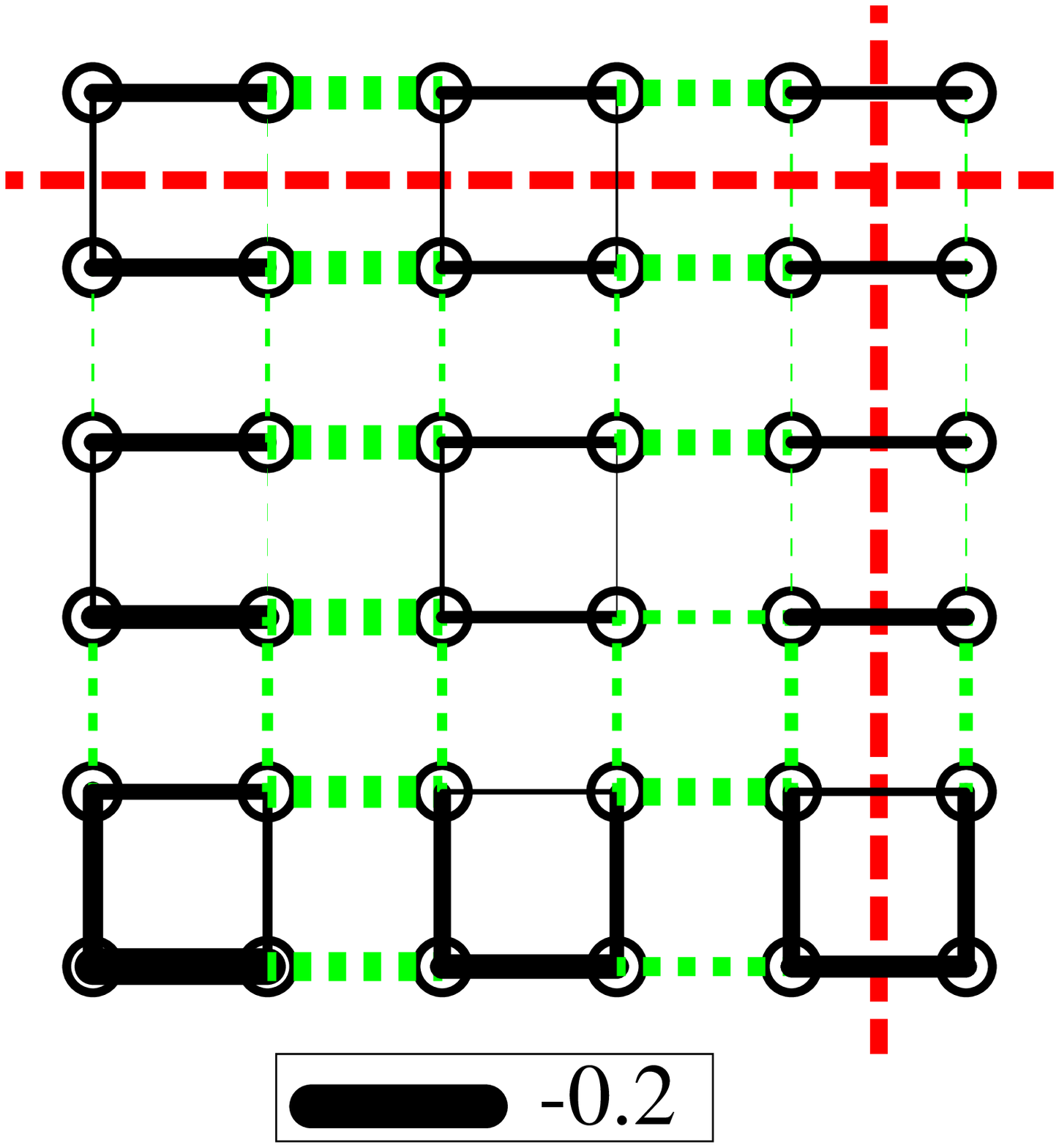,width=6cm}\vspace{0.5cm}\\
  \epsfig{file=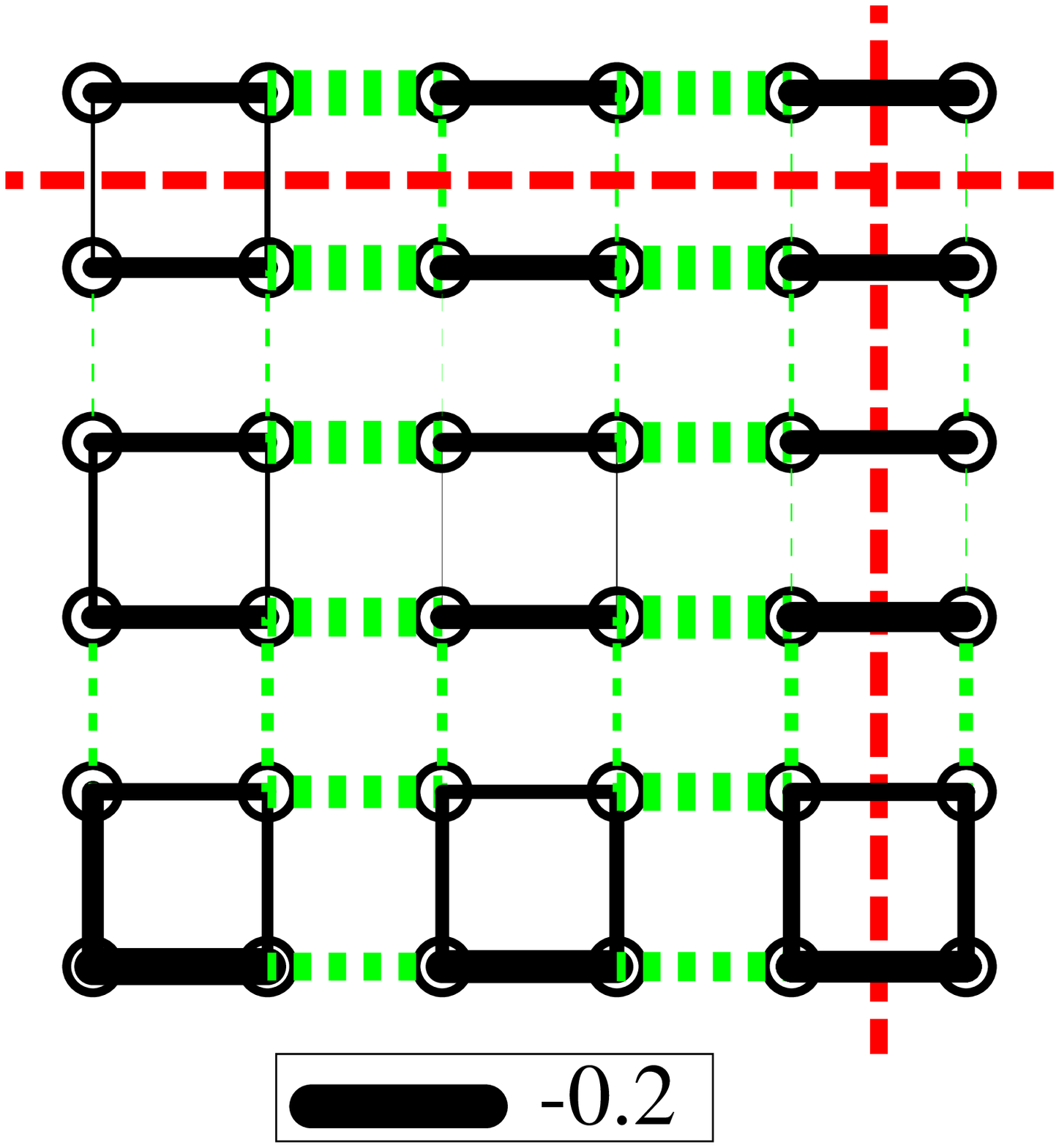,width=6cm}
  \hfill
  \epsfig{file=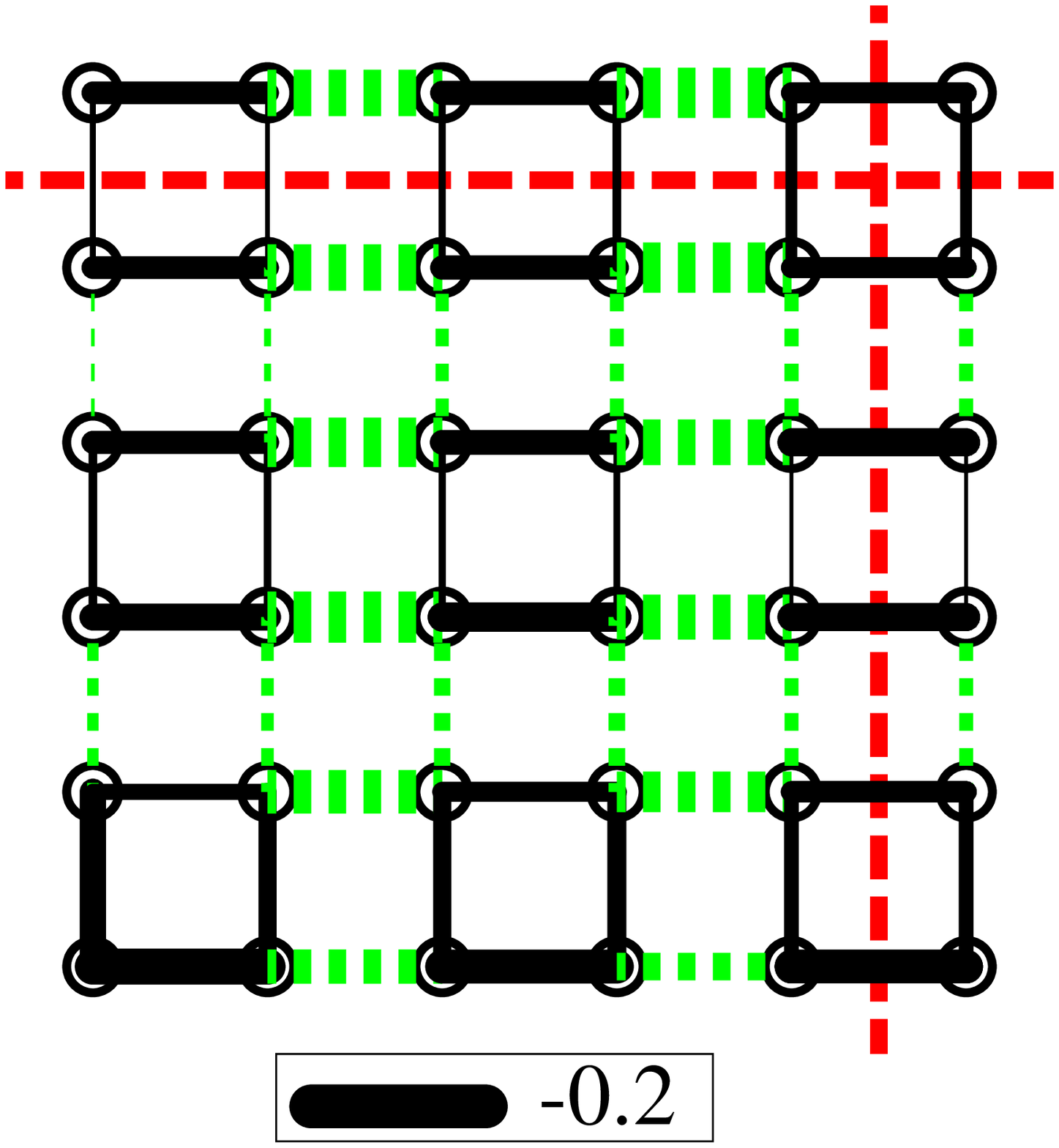,width=6cm}
  \hfill \centering \\*[4mm]
\caption[]{The relative correlation strengths on $10 \times 10$ lattice.
  All other nearest neighbour correlations can be obtained by
  reflection these picture in the two dashed lines. The DMRG guiding
  state follows the meandering sequence of Fig. \ref{fig2}(b). 
  More explanation is given in the
  text.  Reading zig zag from top left to bottom right, the values for $J_2$
  are $J_2=0,\dots,0.5$ in steps of $0.1$. }
\label{fig:variousmeander2}
\end{figure}

\begin{figure}
  \epsfig{file=result.10x10.m75.e0.014.J20.5.meander2.ss.eps,width=6cm}
  \hfill
  \epsfig{file=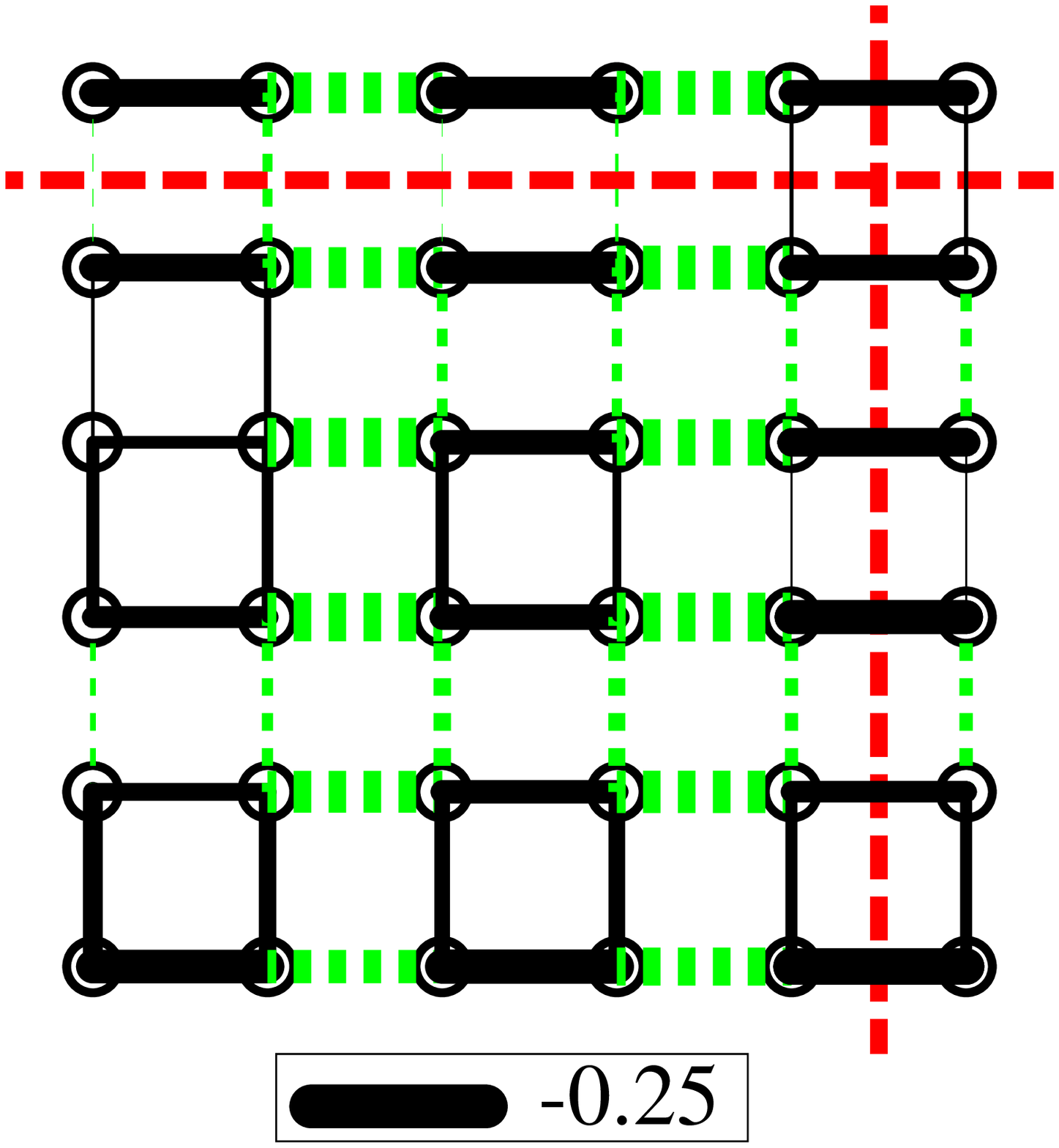,width=6cm} \vspace{0.5cm}\\
  \epsfig{file=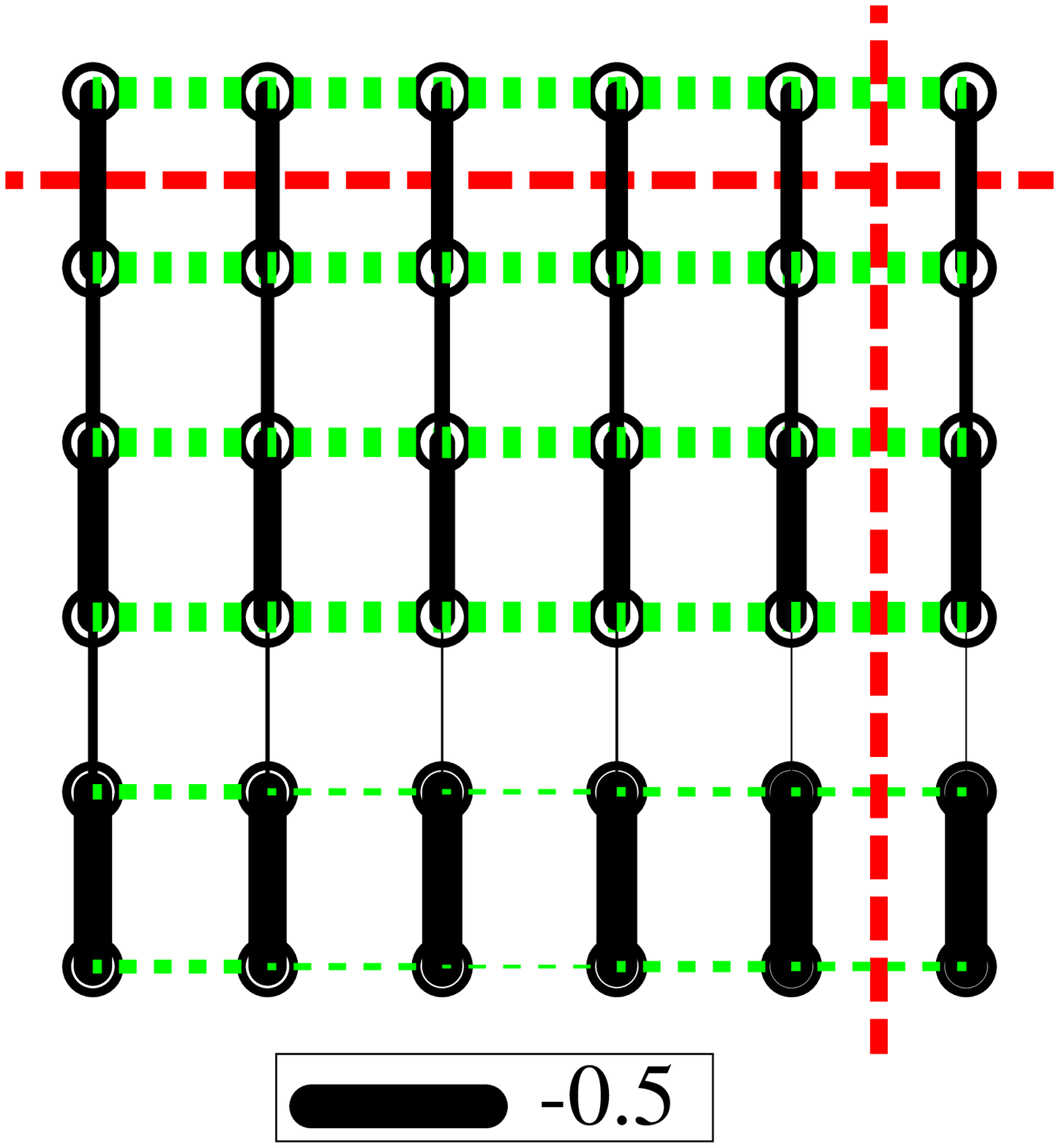,width=6cm}
  \hfill
  \epsfig{file=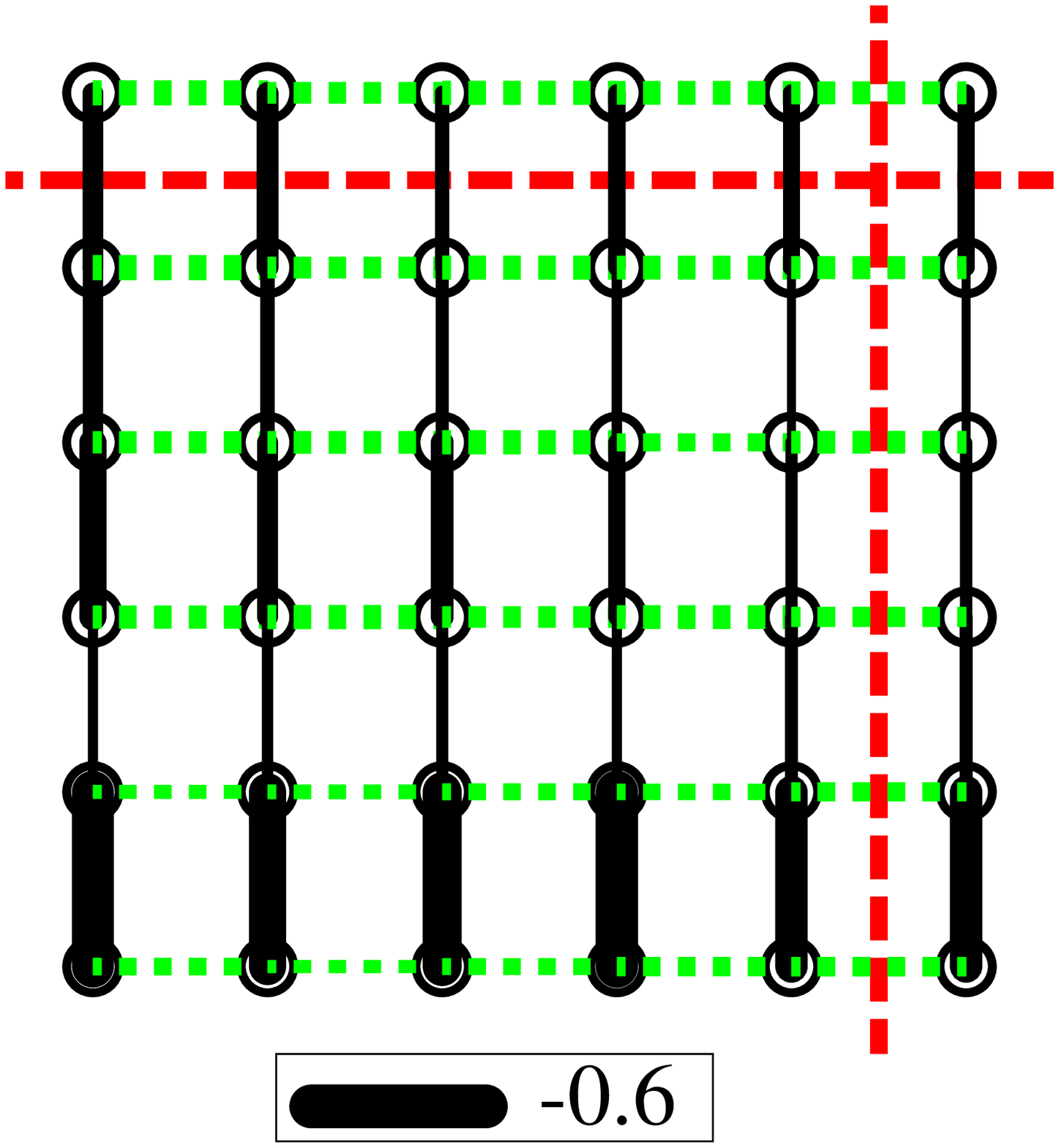,width=6cm}\vspace{0.5cm}\\
  \epsfig{file=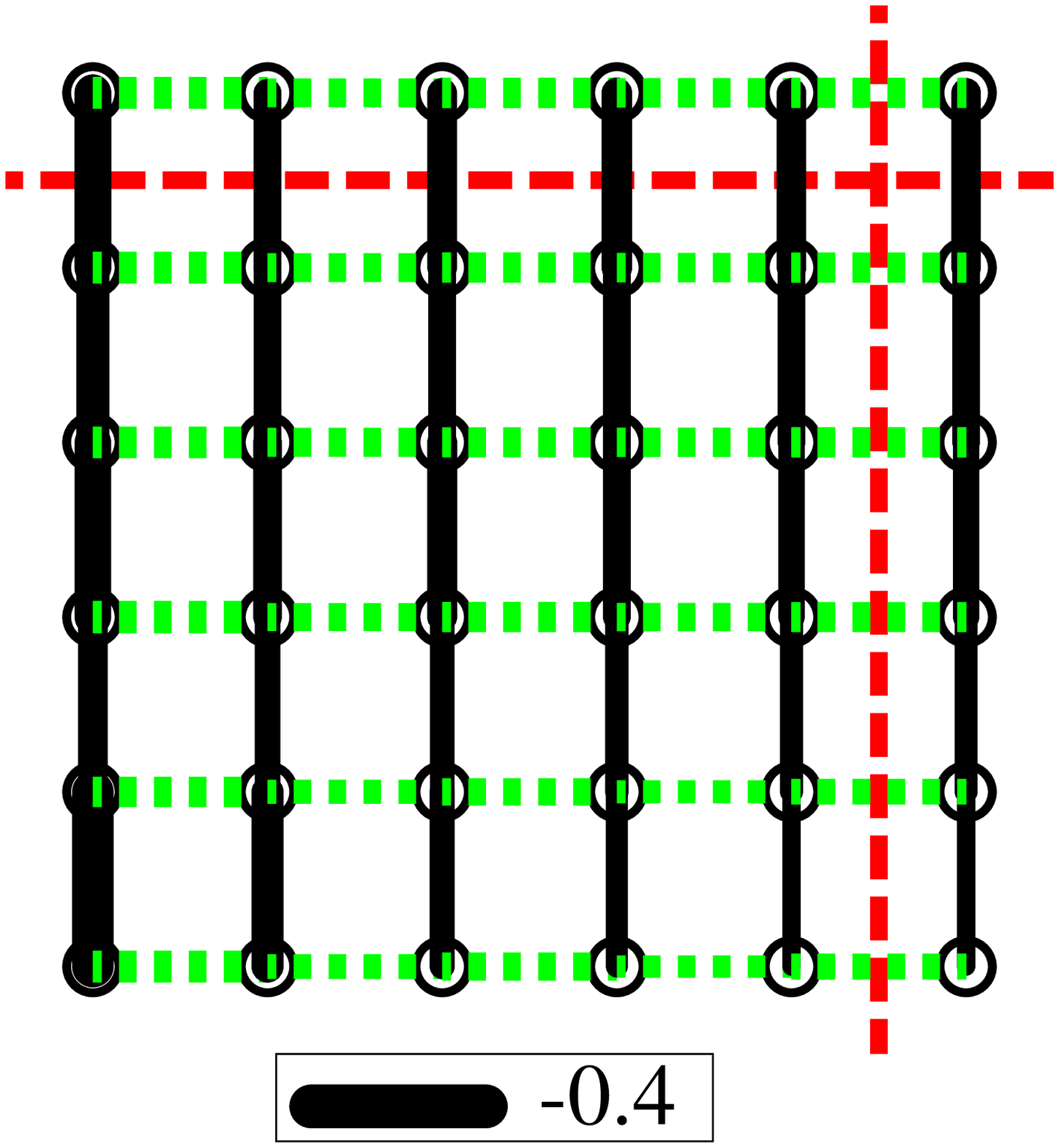,width=6cm}
  \hfill
  \epsfig{file=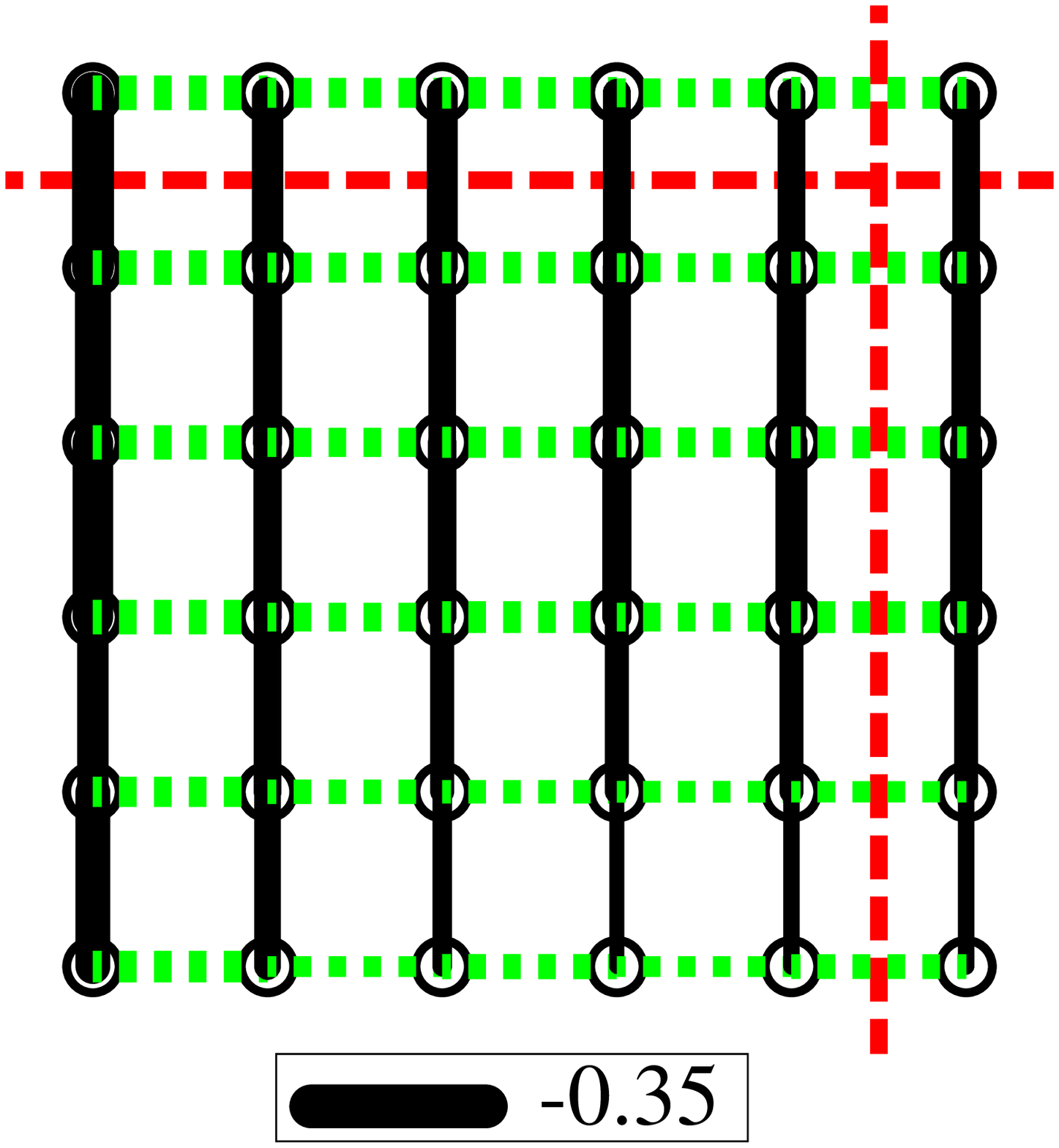,width=6cm}
  \hfill \centering \\*[4mm]
\caption[]{The continuation of figure \ref{fig:variousmeander2}; the
  relative correlation strengths on $10 \times 10$ lattice.
  $J_2=0.5,\dots,1.0$ in steps of $0.1$. }
\label{fig:variousmeander2b}
\end{figure}

\end{document}